\documentclass{iopjournal}

\hypersetup{
    colorlinks,
    linkcolor={red!50!black},
    citecolor={blue!50!black},
    urlcolor={blue!80!black}
}

\usepackage{graphicx,amsfonts,amsmath,amssymb,hyperref,multirow,array}
\usepackage[utf8]{inputenc}
\usepackage[backend=bibtex]{biblatex}

\def\blx@bblfile@bibtex{
  \blx@secinit
  \begingroup
  \blx@bblstart
\begingroup
\makeatletter
\@ifundefined{ver@biblatex.sty}
  {\@latex@error
     {Missing 'biblatex' package}
     {The bibliography requires the 'biblatex' package.}
      \aftergroup\endinput}
  {}
\endgroup

\datalist[entry]{nty/global//global/global}
  \entry{Andreotti13}{book}{}
    \name{author}{3}{}{%
      {{hash=AB}{%
         family={Andreotti},
         familyi={A\bibinitperiod},
         given={B.},
         giveni={B\bibinitperiod},
      }}%
      {{hash=FY}{%
         family={Forterre},
         familyi={F\bibinitperiod},
         given={Y.},
         giveni={Y\bibinitperiod},
      }}%
      {{hash=PO}{%
         family={Pouliquen},
         familyi={P\bibinitperiod},
         given={O.},
         giveni={O\bibinitperiod},
      }}%
    }
    \list{publisher}{1}{%
      {Cambridge University Press}%
    }
    \strng{namehash}{ABFYPO1}
    \strng{fullhash}{ABFYPO1}
    \field{labelnamesource}{author}
    \field{labeltitlesource}{title}
    \field{sortinit}{A}
    \field{sortinithash}{A}
    \verb{doi}
    \verb 10.1017/CBO9781139541008
    \endverb
    \field{title}{Granular Media: Between Fluid and Solid}
    \list{location}{1}{%
      {Cambridge}%
    }
    \field{year}{2013}
  \endentry

  \entry{Artoni21}{article}{}
    \name{author}{4}{}{%
      {{hash=AR}{%
         family={Artoni},
         familyi={A\bibinitperiod},
         given={R.},
         giveni={R\bibinitperiod},
      }}%
      {{hash=LM}{%
         family={Larcher},
         familyi={L\bibinitperiod},
         given={M.},
         giveni={M\bibinitperiod},
      }}%
      {{hash=JJ}{%
         family={Jenkins},
         familyi={J\bibinitperiod},
         given={{J.T.}},
         giveni={J\bibinitperiod},
      }}%
      {{hash=RP}{%
         family={Richard},
         familyi={R\bibinitperiod},
         given={P.},
         giveni={P\bibinitperiod},
      }}%
    }
    \strng{namehash}{AR+1}
    \strng{fullhash}{ARLMJJRP1}
    \field{labelnamesource}{author}
    \field{labeltitlesource}{title}
    \field{sortinit}{A}
    \field{sortinithash}{A}
    \verb{doi}
    \verb 10.1039/D0SM01846E
    \endverb
    \field{issue}{9}
    \field{pages}{2596\bibrangedash 2602}
    \field{title}{Self-diffusion scalings in dense granular flows}
    \field{volume}{17}
    \field{journaltitle}{Soft Matter}
    \field{year}{2021}
  \endentry

  \entry{Balkovsky99}{article}{}
    \name{author}{4}{}{%
      {{hash=BE}{%
         family={Balkovsky},
         familyi={B\bibinitperiod},
         given={E.},
         giveni={E\bibinitperiod},
      }}%
      {{hash=FG}{%
         family={Falkovich},
         familyi={F\bibinitperiod},
         given={G.},
         giveni={G\bibinitperiod},
      }}%
      {{hash=LV}{%
         family={Lebedev},
         familyi={L\bibinitperiod},
         given={V.},
         giveni={V\bibinitperiod},
      }}%
      {{hash=LM}{%
         family={Lysianski},
         familyi={L\bibinitperiod},
         given={M.},
         giveni={M\bibinitperiod},
      }}%
    }
    \strng{namehash}{BE+1}
    \strng{fullhash}{BEFGLVLM1}
    \field{labelnamesource}{author}
    \field{labeltitlesource}{title}
    \field{sortinit}{B}
    \field{sortinithash}{B}
    \verb{doi}
    \verb 10.1063/1.870089
    \endverb
    \field{pages}{2269}
    \field{title}{Large-scale properties of passive scalar advection Available}
    \field{volume}{11}
    \field{journaltitle}{Physics of Fluids}
    \field{year}{1999}
  \endentry

  \entry{Batchelor58}{article}{}
    \name{author}{1}{}{%
      {{hash=BG}{%
         family={Batchelor},
         familyi={B\bibinitperiod},
         given={{G. K.}},
         giveni={G\bibinitperiod},
      }}%
    }
    \strng{namehash}{BG1}
    \strng{fullhash}{BG1}
    \field{labelnamesource}{author}
    \field{labeltitlesource}{title}
    \field{sortinit}{B}
    \field{sortinithash}{B}
    \verb{doi}
    \verb 10.1017/S002211205900009X
    \endverb
    \field{pages}{8}
    \field{title}{Small-scale variation of convected quantities like
  temperature in turbulent fluid Part 1. General discussion and the case of
  small conductivity}
    \field{volume}{5}
    \field{journaltitle}{Fluid Mechanics}
    \field{year}{1958}
  \endentry

  \entry{Batchelor59}{article}{}
    \name{author}{3}{}{%
      {{hash=BG}{%
         family={Batchelor},
         familyi={B\bibinitperiod},
         given={{G. K.}},
         giveni={G\bibinitperiod},
      }}%
      {{hash=HI}{%
         family={Howells},
         familyi={H\bibinitperiod},
         given={{I. D.}},
         giveni={I\bibinitperiod},
      }}%
      {{hash=TA}{%
         family={Townsend},
         familyi={T\bibinitperiod},
         given={{A. A.}},
         giveni={A\bibinitperiod},
      }}%
    }
    \strng{namehash}{BGHITA1}
    \strng{fullhash}{BGHITA1}
    \field{labelnamesource}{author}
    \field{labeltitlesource}{title}
    \field{sortinit}{B}
    \field{sortinithash}{B}
    \verb{doi}
    \verb 10.1017/S0022112059000106
    \endverb
    \field{issue}{1}
    \field{pages}{134\bibrangedash 139}
    \field{title}{Small-scale variation of convected quantities like
  temperature in turbulent fluid Part 2. The case of large conductivity}
    \field{volume}{5}
    \field{journaltitle}{J. Fluid Mech.}
    \field{year}{1959}
  \endentry

  \entry{Canet22}{article}{}
    \name{author}{1}{}{%
      {{hash=CL}{%
         family={Canet},
         familyi={C\bibinitperiod},
         given={L.},
         giveni={L\bibinitperiod},
      }}%
    }
    \strng{namehash}{CL1}
    \strng{fullhash}{CL1}
    \field{labelnamesource}{author}
    \field{labeltitlesource}{title}
    \field{sortinit}{C}
    \field{sortinithash}{C}
    \verb{doi}
    \verb 10.1017/jfm.2022.808
    \endverb
    \field{pages}{1}
    \field{title}{Functional renormalisation group for turbulence}
    \field{volume}{950}
    \field{journaltitle}{J. Fluid. Mech.}
    \field{year}{2022}
  \endentry

  \entry{Canet16}{article}{}
    \name{author}{3}{}{%
      {{hash=CL}{%
         family={Canet},
         familyi={C\bibinitperiod},
         given={L.},
         giveni={L\bibinitperiod},
      }}%
      {{hash=DB}{%
         family={Delamotte},
         familyi={D\bibinitperiod},
         given={B.},
         giveni={B\bibinitperiod},
      }}%
      {{hash=WW}{%
         family={Wschebor},
         familyi={W\bibinitperiod},
         given={W.},
         giveni={W\bibinitperiod},
      }}%
    }
    \strng{namehash}{CLDBWW1}
    \strng{fullhash}{CLDBWW1}
    \field{labelnamesource}{author}
    \field{labeltitlesource}{title}
    \field{sortinit}{C}
    \field{sortinithash}{C}
    \verb{doi}
    \verb 10.1103/PhysRevE.93.063101
    \endverb
    \field{pages}{063101}
    \field{title}{Fully developed isotropic turbulence: Nonperturbative
  renormalization group formalism and fixed-point solution}
    \field{volume}{93}
    \field{journaltitle}{Phys. Rev. E}
    \field{year}{2016}
  \endentry

  \entry{Canet15}{article}{}
    \name{author}{3}{}{%
      {{hash=CL}{%
         family={Canet},
         familyi={C\bibinitperiod},
         given={L.},
         giveni={L\bibinitperiod},
      }}%
      {{hash=DB}{%
         family={Delamotte},
         familyi={D\bibinitperiod},
         given={B.},
         giveni={B\bibinitperiod},
      }}%
      {{hash=WW}{%
         family={Wschebor},
         familyi={W\bibinitperiod},
         given={W.},
         giveni={W\bibinitperiod},
      }}%
    }
    \strng{namehash}{CLDBWW1}
    \strng{fullhash}{CLDBWW1}
    \field{labelnamesource}{author}
    \field{labeltitlesource}{title}
    \field{sortinit}{C}
    \field{sortinithash}{C}
    \verb{doi}
    \verb 10.1103/PhysRevE.91.053004
    \endverb
    \field{pages}{053004}
    \field{title}{Fully developed isotropic turbulence: Symmetries and exact
  identities}
    \field{volume}{91}
    \field{journaltitle}{Phys. Rev. E}
    \field{year}{2015}
  \endentry

  \entry{Canet17}{article}{}
    \name{author}{4}{}{%
      {{hash=CL}{%
         family={Canet},
         familyi={C\bibinitperiod},
         given={L.},
         giveni={L\bibinitperiod},
      }}%
      {{hash=RV}{%
         family={Rossetto},
         familyi={R\bibinitperiod},
         given={V.},
         giveni={V\bibinitperiod},
      }}%
      {{hash=WW}{%
         family={Wschebor},
         familyi={W\bibinitperiod},
         given={W.},
         giveni={W\bibinitperiod},
      }}%
      {{hash=BG}{%
         family={Balarac},
         familyi={B\bibinitperiod},
         given={G.},
         giveni={G\bibinitperiod},
      }}%
    }
    \strng{namehash}{CL+1}
    \strng{fullhash}{CLRVWWBG1}
    \field{labelnamesource}{author}
    \field{labeltitlesource}{title}
    \field{sortinit}{C}
    \field{sortinithash}{C}
    \verb{doi}
    \verb 10.1103/PhysRevE.95.023107
    \endverb
    \field{pages}{023107}
    \field{title}{Spatiotemporal velocity-velocity correlation function in
  fully developed turbulence}
    \field{volume}{95}
    \field{journaltitle}{Phys. Rev. E}
    \field{year}{2017}
  \endentry

  \entry{Cartes22}{article}{}
    \name{author}{4}{}{%
      {{hash=CC}{%
         family={Cartes},
         familyi={C\bibinitperiod},
         given={C.},
         giveni={C\bibinitperiod},
      }}%
      {{hash=TE}{%
         family={Tirapegui},
         familyi={T\bibinitperiod},
         given={E.},
         giveni={E\bibinitperiod},
      }}%
      {{hash=PR}{%
         family={Prandit},
         familyi={P\bibinitperiod},
         given={R.},
         giveni={R\bibinitperiod},
      }}%
      {{hash=BM}{%
         family={Brachet},
         familyi={B\bibinitperiod},
         given={M.},
         giveni={M\bibinitperiod},
      }}%
    }
    \strng{namehash}{CC+1}
    \strng{fullhash}{CCTEPRBM1}
    \field{labelnamesource}{author}
    \field{labeltitlesource}{title}
    \field{sortinit}{C}
    \field{sortinithash}{C}
    \verb{doi}
    \verb 10.1098/rsta.2021.0090
    \endverb
    \field{pages}{20210090}
    \field{title}{The Galerkin-truncated Burgers equation: crossover from
  inviscid-thermalized to Kardar–Parisi–Zhang scaling}
    \field{volume}{380}
    \field{journaltitle}{Phil. Trans. R. Soc. A}
    \field{year}{2022}
  \endentry

  \entry{Chertov95}{article}{}
    \name{author}{4}{}{%
      {{hash=CM}{%
         family={Chertov},
         familyi={C\bibinitperiod},
         given={M.},
         giveni={M\bibinitperiod},
      }}%
      {{hash=FG}{%
         family={Falkovich},
         familyi={F\bibinitperiod},
         given={G.},
         giveni={G\bibinitperiod},
      }}%
      {{hash=KI}{%
         family={Kolokolov},
         familyi={K\bibinitperiod},
         given={I.},
         giveni={I\bibinitperiod},
      }}%
      {{hash=LV}{%
         family={Lebedev},
         familyi={L\bibinitperiod},
         given={V.},
         giveni={V\bibinitperiod},
      }}%
    }
    \strng{namehash}{CM+1}
    \strng{fullhash}{CMFGKILV1}
    \field{labelnamesource}{author}
    \field{labeltitlesource}{title}
    \field{sortinit}{C}
    \field{sortinithash}{C}
    \verb{doi}
    \verb 10.1103/PhysRevE.52.4924
    \endverb
    \field{pages}{5}
    \field{title}{Normal and anomalous scaling of the fourth-order correlation
  function of a randomly advected passive scalar}
    \field{volume}{52}
    \field{journaltitle}{Phys. Rev. E}
    \field{year}{1995}
  \endentry

  \entry{Combe15}{article}{}
    \name{author}{4}{}{%
      {{hash=CG}{%
         family={Combe},
         familyi={C\bibinitperiod},
         given={G.},
         giveni={G\bibinitperiod},
      }}%
      {{hash=RV}{%
         family={Richefeu},
         familyi={R\bibinitperiod},
         given={V.},
         giveni={V\bibinitperiod},
      }}%
      {{hash=SM}{%
         family={Stasiak},
         familyi={S\bibinitperiod},
         given={M.},
         giveni={M\bibinitperiod},
      }}%
      {{hash=AA}{%
         family={Atman},
         familyi={A\bibinitperiod},
         given={{A. P. F.}},
         giveni={A\bibinitperiod},
      }}%
    }
    \strng{namehash}{CG+1}
    \strng{fullhash}{CGRVSMAA1}
    \field{labelnamesource}{author}
    \field{labeltitlesource}{title}
    \field{sortinit}{C}
    \field{sortinithash}{C}
    \verb{doi}
    \verb 10.1103/PhysRevLett.115.238301
    \endverb
    \field{pages}{238301}
    \field{title}{Experimental Validation of a Nonextensive Scaling Law in
  Confined Granular Media}
    \field{volume}{115}
    \field{journaltitle}{Phys. Rev. Lett.}
    \field{year}{2015}
  \endentry

  \entry{Coquand23}{article}{}
    \name{author}{2}{}{%
      {{hash=CO}{%
         family={Coquand},
         familyi={C\bibinitperiod},
         given={O.},
         giveni={O\bibinitperiod},
      }}%
      {{hash=SM}{%
         family={Sperl},
         familyi={S\bibinitperiod},
         given={M.},
         giveni={M\bibinitperiod},
      }}%
    }
    \strng{namehash}{COSM1}
    \strng{fullhash}{COSM1}
    \field{labelnamesource}{author}
    \field{labeltitlesource}{title}
    \field{sortinit}{C}
    \field{sortinithash}{C}
    \verb{doi}
    \verb 10.1103/PhysRevE.109.034901
    \endverb
    \field{pages}{034901}
    \field{title}{Dynamical yield criterion for granular matter from first
  principles}
    \field{volume}{109}
    \field{journaltitle}{Phys. Rev. E}
    \field{year}{2024}
  \endentry

  \entry{Coquand20f}{article}{}
    \name{author}{3}{}{%
      {{hash=CO}{%
         family={Coquand},
         familyi={C\bibinitperiod},
         given={O.},
         giveni={O\bibinitperiod},
      }}%
      {{hash=SM}{%
         family={Sperl},
         familyi={S\bibinitperiod},
         given={M.},
         giveni={M\bibinitperiod},
      }}%
      {{hash=KWT}{%
         family={Kranz},
         familyi={K\bibinitperiod},
         given={W.\bibnamedelima T.},
         giveni={W\bibinitperiod\bibinitdelim T\bibinitperiod},
      }}%
    }
    \strng{namehash}{COSMKWT1}
    \strng{fullhash}{COSMKWT1}
    \field{labelnamesource}{author}
    \field{labeltitlesource}{title}
    \field{sortinit}{C}
    \field{sortinithash}{C}
    \verb{doi}
    \verb 10.1103/PhysRevE.102.032602
    \endverb
    \field{pages}{032602}
    \field{title}{Integration through transients approach to the
  {$\mu(\mathcal{I})$} rheology}
    \field{volume}{102}
    \field{journaltitle}{Phys. Rev. E}
    \field{year}{2020}
  \endentry

  \entry{Coquand24}{misc}{}
    \name{author}{1}{}{%
      {{hash=CO}{%
         family={Coquand},
         familyi={C\bibinitperiod},
         given={Olivier},
         giveni={O\bibinitperiod},
      }}%
    }
    \strng{namehash}{CO1}
    \strng{fullhash}{CO1}
    \field{labelnamesource}{author}
    \field{labeltitlesource}{title}
    \field{sortinit}{C}
    \field{sortinithash}{C}
    \verb{doi}
    \verb 10.48550/arXiv.2408.17353
    \endverb
    \verb{eprint}
    \verb arXiv:2408.17353
    \endverb
    \field{title}{The Liquid-Gas Transition in Granular Matter : a Question of
  Effective Friction?}
    \field{eprintclass}{cond-mat.soft}
    \field{year}{2024}
  \endentry

  \entry{Coquand20g}{misc}{}
    \name{author}{3}{}{%
      {{hash=CO}{%
         family={Coquand},
         familyi={C\bibinitperiod},
         given={Olivier},
         giveni={O\bibinitperiod},
      }}%
      {{hash=KWT}{%
         family={Kranz},
         familyi={K\bibinitperiod},
         given={Wolf\bibnamedelima Till},
         giveni={W\bibinitperiod\bibinitdelim T\bibinitperiod},
      }}%
      {{hash=SM}{%
         family={Sperl},
         familyi={S\bibinitperiod},
         given={Matthias},
         giveni={M\bibinitperiod},
      }}%
    }
    \strng{namehash}{COKWTSM1}
    \strng{fullhash}{COKWTSM1}
    \field{labelnamesource}{author}
    \field{labeltitlesource}{title}
    \field{sortinit}{C}
    \field{sortinithash}{C}
    \verb{doi}
    \verb 10.48550/arXiv.2008.05931
    \endverb
    \verb{eprint}
    \verb arXiv:2008.05931
    \endverb
    \field{title}{Granular rheology: a tale of three time scales}
    \field{eprintclass}{cond-mat.soft}
    \field{year}{2020}
  \endentry

  \entry{Coquand21}{article}{}
    \name{author}{2}{}{%
      {{hash=CO}{%
         family={Coquand},
         familyi={C\bibinitperiod},
         given={Olivier},
         giveni={O\bibinitperiod},
      }}%
      {{hash=SM}{%
         family={Sperl},
         familyi={S\bibinitperiod},
         given={Matthias},
         giveni={M\bibinitperiod},
      }}%
    }
    \strng{namehash}{COSM2}
    \strng{fullhash}{COSM2}
    \field{labelnamesource}{author}
    \field{labeltitlesource}{title}
    \field{sortinit}{C}
    \field{sortinithash}{C}
    \verb{doi}
    \verb 10.1103/PhysRevE.104.014604
    \endverb
    \field{pages}{014604}
    \field{title}{Rheology of granular liquids in extensional flows: Beyond the
  {$\mu(\mathcal{I})$}-law}
    \field{volume}{104}
    \field{journaltitle}{Phys. Rev. E}
    \field{year}{2021}
  \endentry

  \entry{DaCruz05}{article}{}
    \name{author}{5}{}{%
      {{hash=dF}{%
         family={{da Cruz}},
         familyi={d\bibinitperiod},
         given={F.},
         giveni={F\bibinitperiod},
      }}%
      {{hash=ES}{%
         family={Emam},
         familyi={E\bibinitperiod},
         given={S.},
         giveni={S\bibinitperiod},
      }}%
      {{hash=PM}{%
         family={Prochnow},
         familyi={P\bibinitperiod},
         given={M.},
         giveni={M\bibinitperiod},
      }}%
      {{hash=RJN}{%
         family={Roux},
         familyi={R\bibinitperiod},
         given={J.-N.},
         giveni={J\bibinithyphendelim N\bibinitperiod},
      }}%
      {{hash=CF}{%
         family={Chevoir},
         familyi={C\bibinitperiod},
         given={F.},
         giveni={F\bibinitperiod},
      }}%
    }
    \strng{namehash}{dF+1}
    \strng{fullhash}{dFESPMRJNCF1}
    \field{labelnamesource}{author}
    \field{labeltitlesource}{title}
    \field{sortinit}{D}
    \field{sortinithash}{D}
    \verb{doi}
    \verb 10.1103/PhysRevE.72.021309
    \endverb
    \field{pages}{021309}
    \field{title}{Rheophysics of dense granular materials: Discrete simulation
  of plane shear flows}
    \field{volume}{72}
    \field{journaltitle}{Phys. Rev. E}
    \field{year}{2005}
  \endentry

  \entry{Angelo25}{article}{}
    \name{author}{3}{}{%
      {{hash=DO}{%
         family={{D'Angelo}},
         familyi={D\bibinitperiod},
         given={Olfa},
         giveni={O\bibinitperiod},
      }}%
      {{hash=SM}{%
         family={Sperl},
         familyi={S\bibinitperiod},
         given={Matthias},
         giveni={M\bibinitperiod},
      }}%
      {{hash=KWT}{%
         family={Kranz},
         familyi={K\bibinitperiod},
         given={W.\bibnamedelima Till},
         giveni={W\bibinitperiod\bibinitdelim T\bibinitperiod},
      }}%
    }
    \strng{namehash}{DOSMKWT1}
    \strng{fullhash}{DOSMKWT1}
    \field{labelnamesource}{author}
    \field{labeltitlesource}{title}
    \field{sortinit}{D}
    \field{sortinithash}{D}
    \verb{doi}
    \verb 10.1103/PhysRevLett.134.148202
    \endverb
    \field{pages}{148202}
    \field{title}{Rheological Regimes in Agitated Granular Media under Shear}
    \field{volume}{134}
    \field{journaltitle}{Phys. Rev. Lett.}
    \field{year}{2025}
  \endentry

  \entry{Angelo23}{misc}{}
    \name{author}{4}{}{%
      {{hash=DO}{%
         family={{D'Angelo}},
         familyi={D\bibinitperiod},
         given={Olfa},
         giveni={O\bibinitperiod},
      }}%
      {{hash=SA}{%
         family={Shetty},
         familyi={S\bibinitperiod},
         given={Abhishek},
         giveni={A\bibinitperiod},
      }}%
      {{hash=SM}{%
         family={Sperl},
         familyi={S\bibinitperiod},
         given={Matthias},
         giveni={M\bibinitperiod},
      }}%
      {{hash=KWT}{%
         family={Kranz},
         familyi={K\bibinitperiod},
         given={W.\bibnamedelima Till},
         giveni={W\bibinitperiod\bibinitdelim T\bibinitperiod},
      }}%
    }
    \strng{namehash}{DO+1}
    \strng{fullhash}{DOSASMKWT1}
    \field{labelnamesource}{author}
    \field{labeltitlesource}{title}
    \field{sortinit}{D}
    \field{sortinithash}{D}
    \verb{doi}
    \verb 10.48550/arXiv.2309.00413
    \endverb
    \verb{eprint}
    \verb 2309.00413
    \endverb
    \field{title}{The manifold rheology of fluidized granular media}
    \field{eprinttype}{arXiv}
    \field{eprintclass}{cond-mat.soft}
    \field{year}{2023}
  \endentry

  \entry{Betchov57}{article}{}
    \name{author}{1}{}{%
      {{hash=Be}{%
         family={{Betchov}},
         familyi={B\bibinitperiod},
         given={R},
         giveni={R\bibinitperiod},
      }}%
    }
    \strng{namehash}{DOSMKWT1}
    \strng{fullhash}{DOSMKWT1}
    \field{labelnamesource}{author}
    \field{labeltitlesource}{title}
    \field{sortinit}{B}
    \field{sortinithash}{B}
    \verb{doi}
    \verb 10.1017/S0022112057000579
    \endverb
    \field{pages}{205-216}
    \field{title}{On the fine structure of turbulent flows}
    \field{volume}{3}
    \field{journaltitle}{Journal of Fluid Mechanics}
    \field{year}{1957}
  \endentry

  \entry{Dominicis76}{article}{}
    \name{author}{1}{}{%
      {{hash=dC}{%
         family={{de Dominicis}},
         familyi={d\bibinitperiod},
         given={C.},
         giveni={C\bibinitperiod},
      }}%
    }
    \strng{namehash}{dC1}
    \strng{fullhash}{dC1}
    \field{labelnamesource}{author}
    \field{labeltitlesource}{title}
    \field{sortinit}{D}
    \field{sortinithash}{D}
    \verb{doi}
    \verb 10.1051/jphyscol:1976138
    \endverb
    \field{issue}{(C1)}
    \field{pages}{247\bibrangedash 253}
    \field{title}{Techniques de renormalisation de la théorie des champs et
  dynamique des phénomènes critiques}
    \field{volume}{37}
    \field{journaltitle}{J. Phys. {(Paris)} Colloq.}
    \field{year}{1976}
  \endentry

  \entry{DeGiuli16}{article}{}
    \name{author}{3}{}{%
      {{hash=DE}{%
         family={DeGiuli},
         familyi={D\bibinitperiod},
         given={E.},
         giveni={E\bibinitperiod},
      }}%
      {{hash=MJN}{%
         family={McElwaine},
         familyi={M\bibinitperiod},
         given={J.\bibnamedelima N.},
         giveni={J\bibinitperiod\bibinitdelim N\bibinitperiod},
      }}%
      {{hash=WM}{%
         family={Wyart},
         familyi={W\bibinitperiod},
         given={M.},
         giveni={M\bibinitperiod},
      }}%
    }
    \strng{namehash}{DEMJNWM1}
    \strng{fullhash}{DEMJNWM1}
    \field{labelnamesource}{author}
    \field{labeltitlesource}{title}
    \field{sortinit}{D}
    \field{sortinithash}{D}
    \verb{doi}
    \verb 10.1103/PhysRevE.94.012904
    \endverb
    \field{pages}{012904}
    \field{title}{Phase diagram for inertial granular flows}
    \field{volume}{94}
    \field{journaltitle}{Phys. Rev. E}
    \field{year}{2016}
  \endentry

  \entry{DeGiuli17a}{article}{}
    \name{author}{2}{}{%
      {{hash=DE}{%
         family={DeGiuli},
         familyi={D\bibinitperiod},
         given={E.},
         giveni={E\bibinitperiod},
      }}%
      {{hash=WM}{%
         family={Wyart},
         familyi={W\bibinitperiod},
         given={M.},
         giveni={M\bibinitperiod},
      }}%
    }
    \strng{namehash}{DEWM1}
    \strng{fullhash}{DEWM1}
    \field{labelnamesource}{author}
    \field{labeltitlesource}{title}
    \field{sortinit}{D}
    \field{sortinithash}{D}
    \verb{doi}
    \verb 10.1051/epjconf/201714001003
    \endverb
    \field{pages}{01003}
    \field{title}{Unifying Suspension and Granular flows near Jamming}
    \field{volume}{140}
    \field{journaltitle}{Powders and Grains}
    \field{year}{2017}
  \endentry

  \entry{DeGiuli15}{article}{}
    \name{author}{4}{}{%
      {{hash=DE}{%
         family={DeGiuli},
         familyi={D\bibinitperiod},
         given={E.},
         giveni={E\bibinitperiod},
      }}%
      {{hash=DG}{%
         family={D{\"u}ring},
         familyi={D\bibinitperiod},
         given={G.},
         giveni={G\bibinitperiod},
      }}%
      {{hash=LE}{%
         family={Lerner},
         familyi={L\bibinitperiod},
         given={E.},
         giveni={E\bibinitperiod},
      }}%
      {{hash=WM}{%
         family={Wyart},
         familyi={W\bibinitperiod},
         given={M.},
         giveni={M\bibinitperiod},
      }}%
    }
    \strng{namehash}{DE+1}
    \strng{fullhash}{DEDGLEWM1}
    \field{labelnamesource}{author}
    \field{labeltitlesource}{title}
    \field{sortinit}{D}
    \field{sortinithash}{D}
    \verb{doi}
    \verb 10.1103/PhysRevE.91.062206
    \endverb
    \field{pages}{062206}
    \field{title}{Unified theory of inertial granular flows and non-Brownian
  suspensions}
    \field{volume}{91}
    \field{journaltitle}{Phys. Rev. E}
    \field{year}{2015}
  \endentry

  \entry{Eyink06}{article}{}
    \name{author}{2}{}{%
      {{hash=EG}{%
         family={Eyink},
         familyi={E\bibinitperiod},
         given={{G.L.}},
         giveni={G\bibinitperiod},
      }}%
      {{hash=SK}{%
         family={Sreenivasan},
         familyi={S\bibinitperiod},
         given={{K. R.}},
         giveni={K\bibinitperiod},
      }}%
    }
    \strng{namehash}{EGSK1}
    \strng{fullhash}{EGSK1}
    \field{labelnamesource}{author}
    \field{labeltitlesource}{title}
    \field{sortinit}{E}
    \field{sortinithash}{E}
    \verb{doi}
    \verb 10.1103/RevModPhys.78.87
    \endverb
    \field{title}{Onsager and the theory of hydrodynamic turbulence}
    \field{volume}{78}
    \field{journaltitle}{Rev. Mod. Phys.}
    \field{year}{2006}
  \endentry

  \entry{Falkovich01}{article}{}
    \name{author}{3}{}{%
      {{hash=FG}{%
         family={Falkovich},
         familyi={F\bibinitperiod},
         given={G.},
         giveni={G\bibinitperiod},
      }}%
      {{hash=GK}{%
         family={Gaw{\c{e}}dzki},
         familyi={G\bibinitperiod},
         given={K.},
         giveni={K\bibinitperiod},
      }}%
      {{hash=VM}{%
         family={Vergassola},
         familyi={V\bibinitperiod},
         given={M.},
         giveni={M\bibinitperiod},
      }}%
    }
    \strng{namehash}{FGGKVM1}
    \strng{fullhash}{FGGKVM1}
    \field{labelnamesource}{author}
    \field{labeltitlesource}{title}
    \field{sortinit}{F}
    \field{sortinithash}{F}
    \verb{doi}
    \verb 10.1103/RevModPhys.73.913
    \endverb
    \field{pages}{913}
    \field{title}{Particles and fields in fluid turbulence}
    \field{volume}{73}
    \field{journaltitle}{Rev. Mod. Phys.}
    \field{year}{2001}
  \endentry

  \entry{Falkovich94}{article}{}
    \name{author}{2}{}{%
      {{hash=FG}{%
         family={Falkovich},
         familyi={F\bibinitperiod},
         given={G.},
         giveni={G\bibinitperiod},
      }}%
      {{hash=LV}{%
         family={Lebedev},
         familyi={L\bibinitperiod},
         given={V.},
         giveni={V\bibinitperiod},
      }}%
    }
    \strng{namehash}{FGLV1}
    \strng{fullhash}{FGLV1}
    \field{labelnamesource}{author}
    \field{labeltitlesource}{title}
    \field{sortinit}{F}
    \field{sortinithash}{F}
    \verb{doi}
    \verb 10.1103/PhysRevE.50.3883
    \endverb
    \field{pages}{5}
    \field{title}{Universal direct cascade in two-dimensional turbulence}
    \field{volume}{50}
    \field{journaltitle}{Phys. Rev. E}
    \field{year}{1994}
  \endentry

  \entry{Falkovich11}{article}{}
    \name{author}{3}{}{%
      {{hash=FG}{%
         family={Falkovich},
         familyi={F\bibinitperiod},
         given={G.},
         giveni={G\bibinitperiod},
      }}%
      {{hash=LV}{%
         family={Lebedev},
         familyi={L\bibinitperiod},
         given={V.},
         giveni={V\bibinitperiod},
      }}%
      {{hash=SM}{%
         family={Stepanov},
         familyi={S\bibinitperiod},
         given={M.},
         giveni={M\bibinitperiod},
      }}%
    }
    \strng{namehash}{FGLVSM1}
    \strng{fullhash}{FGLVSM1}
    \field{labelnamesource}{author}
    \field{labeltitlesource}{title}
    \field{sortinit}{F}
    \field{sortinithash}{F}
    \verb{doi}
    \verb 10.1103/PhysRevE.83.045301
    \endverb
    \field{pages}{045301(R)}
    \field{title}{Vorticity statistics in the direct cascade of two-dimensional
  turbulence}
    \field{volume}{83}
    \field{journaltitle}{Phys. Rev. E}
    \field{year}{2011}
  \endentry

  \entry{Fedorenko13}{article}{}
    \name{author}{3}{}{%
      {{hash=FA}{%
         family={Fedorenko},
         familyi={F\bibinitperiod},
         given={{A.A.}},
         giveni={A\bibinitperiod},
      }}%
      {{hash=LP}{%
         family={{Le Doussal}},
         familyi={L\bibinitperiod},
         given={P.},
         giveni={P\bibinitperiod},
      }}%
      {{hash=WK}{%
         family={Wiese},
         familyi={W\bibinitperiod},
         given={{K. J.}},
         giveni={K\bibinitperiod},
      }}%
    }
    \strng{namehash}{FALPWK1}
    \strng{fullhash}{FALPWK1}
    \field{labelnamesource}{author}
    \field{labeltitlesource}{title}
    \field{sortinit}{F}
    \field{sortinithash}{F}
    \verb{doi}
    \verb 10.1088/1742-5468/2013/04/P04014
    \endverb
    \field{pages}{P04014}
    \field{title}{Functional renormalization-group approach to decaying
  turbulence}
    \field{journaltitle}{J. Stat. Mech.}
    \field{year}{2013}
  \endentry

  \entry{Forster77}{article}{}
    \name{author}{3}{}{%
      {{hash=FD}{%
         family={Forster},
         familyi={F\bibinitperiod},
         given={D.},
         giveni={D\bibinitperiod},
      }}%
      {{hash=ND}{%
         family={Nelson},
         familyi={N\bibinitperiod},
         given={{D. R.}},
         giveni={D\bibinitperiod},
      }}%
      {{hash=SM}{%
         family={Stephen},
         familyi={S\bibinitperiod},
         given={{M. J.}},
         giveni={M\bibinitperiod},
      }}%
    }
    \strng{namehash}{FDNDSM1}
    \strng{fullhash}{FDNDSM1}
    \field{labelnamesource}{author}
    \field{labeltitlesource}{title}
    \field{sortinit}{F}
    \field{sortinithash}{F}
    \verb{doi}
    \verb 10.1103/PhysRevA.16.732
    \endverb
    \field{pages}{732}
    \field{title}{Large-distance and long-time properties of a randomly stirred
  fluid}
    \field{volume}{16}
    \field{journaltitle}{Phys. Rev. A}
    \field{year}{1977}
  \endentry

  \entry{Forterre18}{article}{}
    \name{author}{2}{}{%
      {{hash=FY}{%
         family={Forterre},
         familyi={F\bibinitperiod},
         given={Y.},
         giveni={Y\bibinitperiod},
      }}%
      {{hash=PO}{%
         family={Pouliquen},
         familyi={P\bibinitperiod},
         given={O.},
         giveni={O\bibinitperiod},
      }}%
    }
    \strng{namehash}{FYPO1}
    \strng{fullhash}{FYPO1}
    \field{labelnamesource}{author}
    \field{labeltitlesource}{title}
    \field{sortinit}{F}
    \field{sortinithash}{F}
    \verb{doi}
    \verb 10.1016/j.crhy.2018.10.003
    \endverb
    \field{pages}{271\bibrangedash 284}
    \field{title}{Physics of particulate flows: From sand avalanche to active
  suspensions in plants}
    \field{volume}{19}
    \field{journaltitle}{C. R. Physique}
    \field{year}{2018}
  \endentry

  \entry{Fuchs09}{article}{}
    \name{author}{2}{}{%
      {{hash=FM}{%
         family={Fuchs},
         familyi={F\bibinitperiod},
         given={M.},
         giveni={M\bibinitperiod},
      }}%
      {{hash=CM}{%
         family={Cates},
         familyi={C\bibinitperiod},
         given={M.},
         giveni={M\bibinitperiod},
      }}%
    }
    \strng{namehash}{FMCM1}
    \strng{fullhash}{FMCM1}
    \field{labelnamesource}{author}
    \field{labeltitlesource}{title}
    \field{sortinit}{F}
    \field{sortinithash}{F}
    \verb{doi}
    \verb 10.1122/1.3119084
    \endverb
    \field{pages}{957\bibrangedash 1000}
    \field{title}{A mode coupling theory for Brownian particles in homogeneous
  steady shear flow}
    \field{volume}{53(4)}
    \field{journaltitle}{J. Rheol.}
    \field{year}{2009}
  \endentry

  \entry{Fuchs03}{article}{}
    \name{author}{2}{}{%
      {{hash=FM}{%
         family={Fuchs},
         familyi={F\bibinitperiod},
         given={M.},
         giveni={M\bibinitperiod},
      }}%
      {{hash=CM}{%
         family={Cates},
         familyi={C\bibinitperiod},
         given={M.},
         giveni={M\bibinitperiod},
      }}%
    }
    \strng{namehash}{FMCM1}
    \strng{fullhash}{FMCM1}
    \field{labelnamesource}{author}
    \field{labeltitlesource}{title}
    \field{sortinit}{F}
    \field{sortinithash}{F}
    \verb{doi}
    \verb 10.1039/B205629A
    \endverb
    \field{pages}{267\bibrangedash 286}
    \field{title}{Schematic models for dynamic yielding of sheared colloidal
  glasses}
    \field{volume}{123}
    \field{journaltitle}{Faraday Discuss.}
    \field{year}{2003}
  \endentry

  \entry{Fuchs02}{article}{}
    \name{author}{2}{}{%
      {{hash=FM}{%
         family={Fuchs},
         familyi={F\bibinitperiod},
         given={M.},
         giveni={M\bibinitperiod},
      }}%
      {{hash=CM}{%
         family={Cates},
         familyi={C\bibinitperiod},
         given={M.},
         giveni={M\bibinitperiod},
      }}%
    }
    \strng{namehash}{FMCM1}
    \strng{fullhash}{FMCM1}
    \field{labelnamesource}{author}
    \field{labeltitlesource}{title}
    \field{sortinit}{F}
    \field{sortinithash}{F}
    \verb{doi}
    \verb 10.1103/PhysRevLett.89.248304
    \endverb
    \field{number}{24}
    \field{pages}{248304}
    \field{title}{Theory of Nonlinear Rheology and Yielding of Dense Colloidal
  Suspensions}
    \field{volume}{89}
    \field{journaltitle}{Phys. Rev. Lett.}
    \field{year}{2002}
  \endentry

  \entry{GDR04}{article}{}
    \name{author}{1}{}{%
      {{hash=G}{%
         family={{GDR MiDi}},
         familyi={G\bibinitperiod},
      }}%
    }
    \strng{namehash}{G1}
    \strng{fullhash}{G1}
    \field{labelnamesource}{author}
    \field{labeltitlesource}{title}
    \field{sortinit}{G}
    \field{sortinithash}{G}
    \verb{doi}
    \verb 10.1140/epje/i2003-10153-0
    \endverb
    \field{pages}{341\bibrangedash 365}
    \field{title}{On dense granular flows}
    \field{volume}{14}
    \field{journaltitle}{Eur. Phys. J. E}
    \field{year}{2004}
  \endentry

  \entry{Gorbunova20}{article}{}
    \name{author}{6}{}{%
      {{hash=GA}{%
         family={Gorbunova},
         familyi={G\bibinitperiod},
         given={A.},
         giveni={A\bibinitperiod},
      }}%
      {{hash=BG}{%
         family={Balarac},
         familyi={B\bibinitperiod},
         given={G.},
         giveni={G\bibinitperiod},
      }}%
      {{hash=BM}{%
         family={Bourgoin},
         familyi={B\bibinitperiod},
         given={M.},
         giveni={M\bibinitperiod},
      }}%
      {{hash=CL}{%
         family={Canet},
         familyi={C\bibinitperiod},
         given={L.},
         giveni={L\bibinitperiod},
      }}%
      {{hash=MN}{%
         family={Mordant},
         familyi={M\bibinitperiod},
         given={N.},
         giveni={N\bibinitperiod},
      }}%
      {{hash=RV}{%
         family={Rossetto},
         familyi={R\bibinitperiod},
         given={V.},
         giveni={V\bibinitperiod},
      }}%
    }
    \strng{namehash}{GA+1}
    \strng{fullhash}{GABGBMCLMNRV1}
    \field{labelnamesource}{author}
    \field{labeltitlesource}{title}
    \field{sortinit}{G}
    \field{sortinithash}{G}
    \verb{doi}
    \verb 10.1103/PhysRevFluids.5.044604
    \endverb
    \field{issue}{4}
    \field{pages}{044604}
    \field{title}{Analysis of the dissipative range of the energy spectrum in
  grid turbulence and in direct numerical simulations}
    \field{volume}{5}
    \field{journaltitle}{Phys. Rev. Fluids}
    \field{year}{2020}
  \endentry

  \entry{Gorbunova21a}{article}{}
    \name{author}{5}{}{%
      {{hash=GA}{%
         family={Gorbunova},
         familyi={G\bibinitperiod},
         given={A.},
         giveni={A\bibinitperiod},
      }}%
      {{hash=BG}{%
         family={Balarac},
         familyi={B\bibinitperiod},
         given={G.},
         giveni={G\bibinitperiod},
      }}%
      {{hash=CL}{%
         family={Canet},
         familyi={C\bibinitperiod},
         given={L.},
         giveni={L\bibinitperiod},
      }}%
      {{hash=EG}{%
         family={Eyink},
         familyi={E\bibinitperiod},
         given={G.},
         giveni={G\bibinitperiod},
      }}%
      {{hash=RV}{%
         family={Rossetto},
         familyi={R\bibinitperiod},
         given={V.},
         giveni={V\bibinitperiod},
      }}%
    }
    \strng{namehash}{GA+1}
    \strng{fullhash}{GABGCLEGRV1}
    \field{labelnamesource}{author}
    \field{labeltitlesource}{title}
    \field{sortinit}{G}
    \field{sortinithash}{G}
    \verb{doi}
    \verb 10.1063/5.0046677
    \endverb
    \field{pages}{055114}
    \field{title}{Spatio-temporal correlations in three-dimensional homogeneous
  and isotropic turbulence}
    \field{volume}{33}
    \field{journaltitle}{Physics of Fluids}
    \field{year}{2021}
  \endentry

  \entry{Goetze08}{book}{}
    \name{author}{1}{}{%
      {{hash=GW}{%
         family={G{\"o}tze},
         familyi={G\bibinitperiod},
         given={W.},
         giveni={W\bibinitperiod},
      }}%
    }
    \list{publisher}{1}{%
      {Oxford University Press}%
    }
    \strng{namehash}{GW1}
    \strng{fullhash}{GW1}
    \field{labelnamesource}{author}
    \field{labeltitlesource}{title}
    \field{sortinit}{G}
    \field{sortinithash}{G}
    \field{isbn}{9780199235346}
    \field{title}{Complex dynamics of Glass-forming liquids}
    \list{location}{1}{%
      {Oxford}%
    }
    \field{year}{2008}
  \endentry

  \entry{Iroshnikov64}{article}{}
    \name{author}{1}{}{%
      {{hash=IP}{%
         family={Iroshnikov},
         familyi={I\bibinitperiod},
         given={{P.S.}},
         giveni={P\bibinitperiod},
      }}%
    }
    \strng{namehash}{IP1}
    \strng{fullhash}{IP1}
    \field{labelnamesource}{author}
    \field{sortinit}{I}
    \field{sortinithash}{I}
    \field{number}{4}
    \field{pages}{566}
    \field{volume}{7}
    \field{journaltitle}{Soviet Astronomy - AJ}
    \field{year}{1964}
  \endentry

  \entry{Janssen76}{article}{}
    \name{author}{1}{}{%
      {{hash=JH}{%
         family={Janssen},
         familyi={J\bibinitperiod},
         given={{H.-K.}},
         giveni={H\bibinitperiod},
      }}%
    }
    \strng{namehash}{JH1}
    \strng{fullhash}{JH1}
    \field{labelnamesource}{author}
    \field{labeltitlesource}{title}
    \field{sortinit}{J}
    \field{sortinithash}{J}
    \verb{doi}
    \verb 10.1007/BF01316547
    \endverb
    \field{pages}{377\bibrangedash 380}
    \field{title}{On a Lagrangean for Classical Field Dynamics and
  Renormalization Group Calculations of Dynamical Critical Properties}
    \field{volume}{23}
    \field{journaltitle}{Z. Phys. B}
    \field{year}{1976}
  \endentry

  \entry{Kelfoun09}{article}{}
    \name{author}{4}{}{%
      {{hash=KK}{%
         family={Kelfoun},
         familyi={K\bibinitperiod},
         given={K.},
         giveni={K\bibinitperiod},
      }}%
      {{hash=SP}{%
         family={Samaniego},
         familyi={S\bibinitperiod},
         given={P.},
         giveni={P\bibinitperiod},
      }}%
      {{hash=PP}{%
         family={Palacios},
         familyi={P\bibinitperiod},
         given={P.},
         giveni={P\bibinitperiod},
      }}%
      {{hash=BD}{%
         family={Barba},
         familyi={B\bibinitperiod},
         given={D.},
         giveni={D\bibinitperiod},
      }}%
    }
    \strng{namehash}{KK+1}
    \strng{fullhash}{KKSPPPBD1}
    \field{labelnamesource}{author}
    \field{labeltitlesource}{title}
    \field{sortinit}{K}
    \field{sortinithash}{K}
    \verb{doi}
    \verb 10.1007/s00445-009-0286-6
    \endverb
    \field{pages}{1057\bibrangedash 1075}
    \field{title}{Testing the suitability of frictional behaviour for
  pyroclastic flow simulation with a well-constrained eruption at Tungurahua
  volcano (Ecuador)}
    \field{volume}{71}
    \field{journaltitle}{Bull. Volcanol.}
    \field{year}{2009}
  \endentry

  \entry{Kharel17}{article}{}
    \name{author}{2}{}{%
      {{hash=KP}{%
         family={Kharel},
         familyi={K\bibinitperiod},
         given={P.},
         giveni={P\bibinitperiod},
      }}%
      {{hash=RP}{%
         family={Rognon},
         familyi={R\bibinitperiod},
         given={P.},
         giveni={P\bibinitperiod},
      }}%
    }
    \strng{namehash}{KPRP1}
    \strng{fullhash}{KPRP1}
    \field{labelnamesource}{author}
    \field{labeltitlesource}{title}
    \field{sortinit}{K}
    \field{sortinithash}{K}
    \verb{doi}
    \verb 10.1103/PhysRevLett.119.178001
    \endverb
    \field{pages}{178001}
    \field{title}{Vortices Enhance Diffusion in Dense Granular Flows}
    \field{volume}{119}
    \field{journaltitle}{Phys. Rev. Lett.}
    \field{year}{2017}
  \endentry

  \entry{Kolmogorov41c}{article}{}
    \name{author}{1}{}{%
      {{hash=KA}{%
         family={Kolmogorov},
         familyi={K\bibinitperiod},
         given={A.},
         giveni={A\bibinitperiod},
      }}%
    }
    \strng{namehash}{KA1}
    \strng{fullhash}{KA1}
    \field{labelnamesource}{author}
    \field{labeltitlesource}{title}
    \field{sortinit}{K}
    \field{sortinithash}{K}
    \field{pages}{16}
    \field{title}{Dissipation of energy in the locally isotropic Turbulence}
    \field{volume}{32}
    \field{journaltitle}{Dokl. Akad. Nauk SSSR}
    \field{year}{1941}
  \endentry

  \entry{Kolmogorov41b}{article}{}
    \name{author}{1}{}{%
      {{hash=KA}{%
         family={Kolmogorov},
         familyi={K\bibinitperiod},
         given={A.},
         giveni={A\bibinitperiod},
      }}%
    }
    \strng{namehash}{KA1}
    \strng{fullhash}{KA1}
    \field{labelnamesource}{author}
    \field{labeltitlesource}{title}
    \field{sortinit}{K}
    \field{sortinithash}{K}
    \field{pages}{538}
    \field{title}{On the degeneration of isotropic turbulence in an
  incompressible viscous fluid}
    \field{volume}{31}
    \field{journaltitle}{Dokl. Akad. Nauk SSSR}
    \field{year}{1941}
  \endentry

  \entry{Kolmogorov41a}{article}{}
    \name{author}{1}{}{%
      {{hash=KA}{%
         family={Kolmogorov},
         familyi={K\bibinitperiod},
         given={A.},
         giveni={A\bibinitperiod},
      }}%
    }
    \strng{namehash}{KA1}
    \strng{fullhash}{KA1}
    \field{labelnamesource}{author}
    \field{labeltitlesource}{title}
    \field{sortinit}{K}
    \field{sortinithash}{K}
    \field{pages}{299}
    \field{title}{The Local Structure of Turbulence in Incompressible Viscous
  Fluid for Very Large Reynolds' Numbers}
    \field{volume}{30}
    \field{journaltitle}{Dokl. Akad. Nauk SSSR}
    \field{year}{1941}
  \endentry

  \entry{Kraichnan65}{article}{}
    \name{author}{1}{}{%
      {{hash=KR}{%
         family={Kraichnan},
         familyi={K\bibinitperiod},
         given={{R. H.}},
         giveni={R\bibinitperiod},
      }}%
    }
    \strng{namehash}{KR1}
    \strng{fullhash}{KR1}
    \field{labelnamesource}{author}
    \field{sortinit}{K}
    \field{sortinithash}{K}
    \field{issue}{7}
    \field{pages}{1385}
    \field{volume}{8}
    \field{journaltitle}{Physics of Fluids}
    \field{year}{1965}
  \endentry

  \entry{Kraichnan74a}{article}{}
    \name{author}{1}{}{%
      {{hash=KR}{%
         family={Kraichnan},
         familyi={K\bibinitperiod},
         given={{R. H.}},
         giveni={R\bibinitperiod},
      }}%
    }
    \strng{namehash}{KR1}
    \strng{fullhash}{KR1}
    \field{labelnamesource}{author}
    \field{labeltitlesource}{title}
    \field{sortinit}{K}
    \field{sortinithash}{K}
    \verb{doi}
    \verb 10.1017/S0022112074001881
    \endverb
    \field{issue}{4}
    \field{pages}{737\bibrangedash 762}
    \field{title}{Convection of a passive scalar by a quasi-uniform random
  straining field}
    \field{volume}{64}
    \field{journaltitle}{J. Fluid Mech.}
    \field{year}{1974}
  \endentry

  \entry{Kraichnan71}{article}{}
    \name{author}{1}{}{%
      {{hash=KR}{%
         family={Kraichnan},
         familyi={K\bibinitperiod},
         given={{R. H.}},
         giveni={R\bibinitperiod},
      }}%
    }
    \strng{namehash}{KR1}
    \strng{fullhash}{KR1}
    \field{labelnamesource}{author}
    \field{labeltitlesource}{title}
    \field{sortinit}{K}
    \field{sortinithash}{K}
    \verb{doi}
    \verb 10.1017/S0022112071001216
    \endverb
    \field{issue}{3}
    \field{pages}{525\bibrangedash 535}
    \field{title}{Inertial-range transfer in two- and three-dimensional
  turbulence}
    \field{volume}{47}
    \field{journaltitle}{J. Fluid Mech.}
    \field{year}{1971}
  \endentry

  \entry{Kraichnan67}{article}{}
    \name{author}{1}{}{%
      {{hash=KR}{%
         family={Kraichnan},
         familyi={K\bibinitperiod},
         given={{R. H.}},
         giveni={R\bibinitperiod},
      }}%
    }
    \strng{namehash}{KR1}
    \strng{fullhash}{KR1}
    \field{labelnamesource}{author}
    \field{labeltitlesource}{title}
    \field{sortinit}{K}
    \field{sortinithash}{K}
    \verb{doi}
    \verb 10.1063/1.1762301
    \endverb
    \field{issue}{7}
    \field{pages}{1417}
    \field{title}{Inertial Ranges in Two‐Dimensional Turbulence}
    \field{volume}{10}
    \field{journaltitle}{Physics of Fluids}
    \field{year}{1967}
  \endentry

  \entry{Kraichnan74b}{article}{}
    \name{author}{1}{}{%
      {{hash=KR}{%
         family={Kraichnan},
         familyi={K\bibinitperiod},
         given={{R. H.}},
         giveni={R\bibinitperiod},
      }}%
    }
    \strng{namehash}{KR1}
    \strng{fullhash}{KR1}
    \field{labelnamesource}{author}
    \field{labeltitlesource}{title}
    \field{sortinit}{K}
    \field{sortinithash}{K}
    \verb{doi}
    \verb 10.1017/S002211207400070X
    \endverb
    \field{issue}{2}
    \field{pages}{305\bibrangedash 330}
    \field{title}{On Kolmogorov's inertial-range theories}
    \field{volume}{62}
    \field{journaltitle}{J. Fluid Mech.}
    \field{year}{1974}
  \endentry

  \entry{Kraichnan75}{article}{}
    \name{author}{1}{}{%
      {{hash=KR}{%
         family={Kraichnan},
         familyi={K\bibinitperiod},
         given={{R. H.}},
         giveni={R\bibinitperiod},
      }}%
    }
    \strng{namehash}{KR1}
    \strng{fullhash}{KR1}
    \field{labelnamesource}{author}
    \field{labeltitlesource}{title}
    \field{sortinit}{K}
    \field{sortinithash}{K}
    \verb{doi}
    \verb 10.1017/S0022112075000225
    \endverb
    \field{issue}{1}
    \field{pages}{155\bibrangedash 175}
    \field{title}{Statistical dynamics of two-dimensional flow}
    \field{volume}{67}
    \field{journaltitle}{J. Fluid Mech.}
    \field{year}{1975}
  \endentry

  \entry{Kranz14}{article}{}
    \name{author}{1}{}{%
      {{hash=KW}{%
         family={Kranz},
         familyi={K\bibinitperiod},
         given={W.T.},
         giveni={W\bibinitperiod},
      }}%
    }
    \strng{namehash}{KW1}
    \strng{fullhash}{KW1}
    \field{labelnamesource}{author}
    \field{labeltitlesource}{title}
    \field{sortinit}{K}
    \field{sortinithash}{K}
    \verb{doi}
    \verb 10.1088/1742-5468/2014/02/P02010
    \endverb
    \field{pages}{P02010}
    \field{title}{A classical long-time tail in a driven granular fluid}
    \field{journaltitle}{J. Stat. Mech.}
    \field{year}{2014}
  \endentry

  \entry{Kranz20}{article}{}
    \name{author}{5}{}{%
      {{hash=KW}{%
         family={Kranz},
         familyi={K\bibinitperiod},
         given={W.T.},
         giveni={W\bibinitperiod},
      }}%
      {{hash=FF}{%
         family={Frahsa},
         familyi={F\bibinitperiod},
         given={F.},
         giveni={F\bibinitperiod},
      }}%
      {{hash=ZA}{%
         family={Zippelius},
         familyi={Z\bibinitperiod},
         given={A.},
         giveni={A\bibinitperiod},
      }}%
      {{hash=FM}{%
         family={Fuchs},
         familyi={F\bibinitperiod},
         given={M.},
         giveni={M\bibinitperiod},
      }}%
      {{hash=SM}{%
         family={Sperl},
         familyi={S\bibinitperiod},
         given={M.},
         giveni={M\bibinitperiod},
      }}%
    }
    \strng{namehash}{KW+1}
    \strng{fullhash}{KWFFZAFMSM1}
    \field{labelnamesource}{author}
    \field{labeltitlesource}{title}
    \field{sortinit}{K}
    \field{sortinithash}{K}
    \verb{doi}
    \verb 10.1103/PhysRevFluids.5.024305
    \endverb
    \field{pages}{024305}
    \field{title}{Integration through transients for inelastic hard sphere
  fluids}
    \field{volume}{5}
    \field{journaltitle}{Phys. Rev. Fluids}
    \field{year}{2020}
  \endentry

  \entry{Kranz18}{article}{}
    \name{author}{5}{}{%
      {{hash=KW}{%
         family={Kranz},
         familyi={K\bibinitperiod},
         given={W.T.},
         giveni={W\bibinitperiod},
      }}%
      {{hash=FF}{%
         family={Frahsa},
         familyi={F\bibinitperiod},
         given={F.},
         giveni={F\bibinitperiod},
      }}%
      {{hash=ZA}{%
         family={Zippelius},
         familyi={Z\bibinitperiod},
         given={A.},
         giveni={A\bibinitperiod},
      }}%
      {{hash=FM}{%
         family={Fuchs},
         familyi={F\bibinitperiod},
         given={M.},
         giveni={M\bibinitperiod},
      }}%
      {{hash=SM}{%
         family={Sperl},
         familyi={S\bibinitperiod},
         given={M.},
         giveni={M\bibinitperiod},
      }}%
    }
    \strng{namehash}{KW+1}
    \strng{fullhash}{KWFFZAFMSM1}
    \field{labelnamesource}{author}
    \field{labeltitlesource}{title}
    \field{sortinit}{K}
    \field{sortinithash}{K}
    \verb{doi}
    \verb 10.1103/PhysRevLett.121.148002
    \endverb
    \field{pages}{148002}
    \field{title}{Rheology of Inelastic Hard Spheres at Finite Density and
  Shear Rate}
    \field{volume}{121}
    \field{journaltitle}{Phys. Rev. Lett.}
    \field{year}{2018}
  \endentry

  \entry{Martin73}{article}{}
    \name{author}{3}{}{%
      {{hash=MP}{%
         family={Martin},
         familyi={M\bibinitperiod},
         given={{P. C.}},
         giveni={P\bibinitperiod},
      }}%
      {{hash=SE}{%
         family={Siggia},
         familyi={S\bibinitperiod},
         given={{E. D. }},
         giveni={E\bibinitperiod},
      }}%
      {{hash=RH}{%
         family={Rose},
         familyi={R\bibinitperiod},
         given={{H. A.}},
         giveni={H\bibinitperiod},
      }}%
    }
    \strng{namehash}{MPSERH1}
    \strng{fullhash}{MPSERH1}
    \field{labelnamesource}{author}
    \field{labeltitlesource}{title}
    \field{sortinit}{M}
    \field{sortinithash}{M}
    \verb{doi}
    \verb 10.1103/PhysRevA.8.423
    \endverb
    \field{pages}{423\bibrangedash 437}
    \field{title}{Statistical Dynamics of Classical Systems}
    \field{volume}{8}
    \field{journaltitle}{Phys. Rev. A}
    \field{year}{1973}
  \endentry

  \entry{Onsager49}{article}{}
    \name{author}{1}{}{%
      {{hash=OL}{%
         family={Onsager},
         familyi={O\bibinitperiod},
         given={L.},
         giveni={L\bibinitperiod},
      }}%
    }
    \strng{namehash}{OL1}
    \strng{fullhash}{OL1}
    \field{labelnamesource}{author}
    \field{labeltitlesource}{title}
    \field{sortinit}{O}
    \field{sortinithash}{O}
    \verb{doi}
    \verb 10.1007/BF02780991
    \endverb
    \field{title}{Statistical Hydrodynamics}
    \field{volume}{2}
    \field{journaltitle}{Supplemento al volume VI serie IX del nuovo cimento}
    \field{year}{1949}
  \endentry

  \entry{Oyama19}{article}{}
    \name{author}{3}{}{%
      {{hash=ON}{%
         family={Oyama},
         familyi={O\bibinitperiod},
         given={N.},
         giveni={N\bibinitperiod},
      }}%
      {{hash=MH}{%
         family={Mizuno},
         familyi={M\bibinitperiod},
         given={H.},
         giveni={H\bibinitperiod},
      }}%
      {{hash=SK}{%
         family={Saitoh},
         familyi={S\bibinitperiod},
         given={K.},
         giveni={K\bibinitperiod},
      }}%
    }
    \strng{namehash}{ONMHSK1}
    \strng{fullhash}{ONMHSK1}
    \field{labelnamesource}{author}
    \field{labeltitlesource}{title}
    \field{sortinit}{O}
    \field{sortinithash}{O}
    \verb{doi}
    \verb 10.1103/PhysRevLett.122.188004
    \endverb
    \field{pages}{188004}
    \field{title}{Avalanche Interpretation of the Power-Law Energy Spectrum in
  Three-Dimensional Dense Granular Flows}
    \field{volume}{122}
    \field{journaltitle}{Phys. Rev. Lett.}
    \field{year}{2019}
  \endentry

  \entry{Prandtl26a}{article}{}
    \name{author}{1}{}{%
      {{hash=PL}{%
         family={Prandtl},
         familyi={P\bibinitperiod},
         given={L.},
         giveni={L\bibinitperiod},
      }}%
    }
    \strng{namehash}{PL1}
    \strng{fullhash}{PL1}
    \field{labelnamesource}{author}
    \field{labeltitlesource}{title}
    \field{sortinit}{P}
    \field{sortinithash}{P}
    \field{pages}{355\bibrangedash 358}
    \field{title}{Aufgaben der Str{\"o}mungsforschung}
    \field{volume}{14}
    \field{journaltitle}{Naturwissenschaften}
    \field{year}{1926}
  \endentry

  \entry{Prandtl26b}{article}{}
    \name{author}{1}{}{%
      {{hash=PL}{%
         family={Prandtl},
         familyi={P\bibinitperiod},
         given={L.},
         giveni={L\bibinitperiod},
      }}%
    }
    \strng{namehash}{PL1}
    \strng{fullhash}{PL1}
    \field{labelnamesource}{author}
    \field{labeltitlesource}{title}
    \field{sortinit}{P}
    \field{sortinithash}{P}
    \field{title}{Bericht {\"u}ber neuere Turbulenzforschung}
    \field{volume}{1}
    \field{journaltitle}{Verein Deutscher Ingenieure Verlag{ ,} Berlin}
    \field{year}{1926}
  \endentry

  \entry{Prandtl26c}{article}{}
    \name{author}{1}{}{%
      {{hash=PL}{%
         family={Prandtl},
         familyi={P\bibinitperiod},
         given={L.},
         giveni={L\bibinitperiod},
      }}%
    }
    \strng{namehash}{PL1}
    \strng{fullhash}{PL1}
    \field{labelnamesource}{author}
    \field{labeltitlesource}{title}
    \field{sortinit}{P}
    \field{sortinithash}{P}
    \field{pages}{62\bibrangedash 74}
    \field{title}{{\"U}ber Die Ausgebildete Turbulenz}
    \field{journaltitle}{Proc. Int. Congr. for Applied Mechanics{ ,}
  Z{\"u}rich}
    \field{year}{1926}
  \endentry

  \entry{Radjai02}{article}{}
    \name{author}{2}{}{%
      {{hash=RF}{%
         family={Radjai},
         familyi={R\bibinitperiod},
         given={F.},
         giveni={F\bibinitperiod},
      }}%
      {{hash=RS}{%
         family={Roux},
         familyi={R\bibinitperiod},
         given={S.},
         giveni={S\bibinitperiod},
      }}%
    }
    \strng{namehash}{RFRS1}
    \strng{fullhash}{RFRS1}
    \field{labelnamesource}{author}
    \field{labeltitlesource}{title}
    \field{sortinit}{R}
    \field{sortinithash}{R}
    \verb{doi}
    \verb 10.1103/PhysRevLett.89.064302
    \endverb
    \field{pages}{064302}
    \field{title}{Turbulentlike Fluctuations in Quasistatic Flow of Granular
  Media}
    \field{volume}{89}
    \field{journaltitle}{Phys. Rev. Lett.}
    \field{year}{2002}
  \endentry

  \entry{Richefeu12}{article}{}
    \name{author}{3}{}{%
      {{hash=RV}{%
         family={Richefeu},
         familyi={R\bibinitperiod},
         given={V.},
         giveni={V\bibinitperiod},
      }}%
      {{hash=CG}{%
         family={Combe},
         familyi={C\bibinitperiod},
         given={G.},
         giveni={G\bibinitperiod},
      }}%
      {{hash=VG}{%
         family={Viggiani},
         familyi={V\bibinitperiod},
         given={G.},
         giveni={G\bibinitperiod},
      }}%
    }
    \strng{namehash}{RVCGVG1}
    \strng{fullhash}{RVCGVG1}
    \field{labelnamesource}{author}
    \field{labeltitlesource}{title}
    \field{sortinit}{R}
    \field{sortinithash}{R}
    \verb{doi}
    \verb 10.1680/geolett.12.00029
    \endverb
    \field{pages}{113\bibrangedash 118}
    \field{title}{An experimental assessment of displacement fluctuations in a
  2D granular material subjected to shear}
    \field{volume}{2}
    \field{journaltitle}{G{\'e}otechnique Letters}
    \field{year}{2012}
  \endentry

  \entry{Saitoh20}{article}{}
    \name{author}{2}{}{%
      {{hash=SK}{%
         family={Saitoh},
         familyi={S\bibinitperiod},
         given={K.},
         giveni={K\bibinitperiod},
      }}%
      {{hash=KT}{%
         family={Kawasaki},
         familyi={K\bibinitperiod},
         given={T.},
         giveni={T\bibinitperiod},
      }}%
    }
    \strng{namehash}{SKKT1}
    \strng{fullhash}{SKKT1}
    \field{labelnamesource}{author}
    \field{labeltitlesource}{title}
    \field{sortinit}{S}
    \field{sortinithash}{S}
    \verb{doi}
    \verb 10.3389/fphy.2020.00099
    \endverb
    \field{pages}{99}
    \field{title}{Critical Scaling of Diffusion Coefficients and Size of Rigid
  Clusters of Soft Athermal Particles Under Shear}
    \field{volume}{8}
    \field{journaltitle}{Frontiers in physics}
    \field{year}{2020}
  \endentry

  \entry{Saitoh16a}{article}{}
    \name{author}{2}{}{%
      {{hash=SK}{%
         family={Saitoh},
         familyi={S\bibinitperiod},
         given={K.},
         giveni={K\bibinitperiod},
      }}%
      {{hash=MH}{%
         family={Mizuno},
         familyi={M\bibinitperiod},
         given={H.},
         giveni={H\bibinitperiod},
      }}%
    }
    \strng{namehash}{SKMH1}
    \strng{fullhash}{SKMH1}
    \field{labelnamesource}{author}
    \field{labeltitlesource}{title}
    \field{sortinit}{S}
    \field{sortinithash}{S}
    \verb{doi}
    \verb 10.1039/C5SM02760H
    \endverb
    \field{pages}{1360\bibrangedash 1367}
    \field{title}{Anomalous energy cascades in dense granular materials
  yielding under simple shear deformations}
    \field{volume}{12}
    \field{journaltitle}{Soft matter}
    \field{year}{2016}
  \endentry

  \entry{Saitoh16}{article}{}
    \name{author}{2}{}{%
      {{hash=SK}{%
         family={Saitoh},
         familyi={S\bibinitperiod},
         given={K.},
         giveni={K\bibinitperiod},
      }}%
      {{hash=MH}{%
         family={Mizuno},
         familyi={M\bibinitperiod},
         given={H.},
         giveni={H\bibinitperiod},
      }}%
    }
    \strng{namehash}{SKMH1}
    \strng{fullhash}{SKMH1}
    \field{labelnamesource}{author}
    \field{labeltitlesource}{title}
    \field{sortinit}{S}
    \field{sortinithash}{S}
    \verb{doi}
    \verb 10.1103/PhysRevE.94.022908
    \endverb
    \field{pages}{022908}
    \field{title}{Enstrophy cascades in two-dimensional dense granular flows}
    \field{volume}{94}
    \field{journaltitle}{Physical Review E}
    \field{year}{2016}
  \endentry

  \entry{Singh23}{article}{}
    \name{author}{2}{}{%
      {{hash=SA}{%
         family={Singh},
         familyi={S\bibinitperiod},
         given={A.},
         giveni={A\bibinitperiod},
      }}%
      {{hash=SK}{%
         family={Saitoh},
         familyi={S\bibinitperiod},
         given={K.},
         giveni={K\bibinitperiod},
      }}%
    }
    \strng{namehash}{SASK1}
    \strng{fullhash}{SASK1}
    \field{labelnamesource}{author}
    \field{labeltitlesource}{title}
    \field{sortinit}{S}
    \field{sortinithash}{S}
    \verb{doi}
    \verb 10.1039/D3SM00510K
    \endverb
    \field{pages}{6631}
    \field{title}{Scaling relationships between viscosity and diffusivity in
  shear-thickening suspensions}
    \field{volume}{19}
    \field{journaltitle}{Soft Matter}
    \field{year}{2023}
  \endentry

  \entry{Tarpin19}{article}{}
    \name{author}{4}{}{%
      {{hash=TM}{%
         family={Tarpin},
         familyi={T\bibinitperiod},
         given={M.},
         giveni={M\bibinitperiod},
      }}%
      {{hash=CL}{%
         family={Canet},
         familyi={C\bibinitperiod},
         given={L.},
         giveni={L\bibinitperiod},
      }}%
      {{hash=PC}{%
         family={Pagani},
         familyi={P\bibinitperiod},
         given={C.},
         giveni={C\bibinitperiod},
      }}%
      {{hash=WN}{%
         family={Wschebor},
         familyi={W\bibinitperiod},
         given={N.},
         giveni={N\bibinitperiod},
      }}%
    }
    \strng{namehash}{TM+1}
    \strng{fullhash}{TMCLPCWN1}
    \field{labelnamesource}{author}
    \field{labeltitlesource}{title}
    \field{sortinit}{T}
    \field{sortinithash}{T}
    \verb{doi}
    \verb 10.1088/1751-8121/aaf3f0
    \endverb
    \field{issue}{8}
    \field{pages}{085501}
    \field{title}{Stationary, isotropic and homogeneous two-dimensional
  turbulence: a first non-perturbative renormalization group approach}
    \field{volume}{52}
    \field{journaltitle}{Journal of Physics A : Mathematical and Theoretical}
    \field{year}{2019}
  \endentry

  \entry{Heisenberg48}{article}{}
    \name{author}{1}{}{%
      {{hash=vW}{%
         family={{von Heisenberg}},
         familyi={v\bibinitperiod},
         given={W.},
         giveni={W\bibinitperiod},
      }}%
    }
    \strng{namehash}{vW1}
    \strng{fullhash}{vW1}
    \field{labelnamesource}{author}
    \field{labeltitlesource}{title}
    \field{sortinit}{V}
    \field{sortinithash}{V}
    \verb{doi}
    \verb 10.1007/BF01668899
    \endverb
    \field{pages}{628\bibrangedash 657}
    \field{title}{Zur statistischen Theorie der Turbulenz}
    \field{volume}{124}
    \field{journaltitle}{Z. Physik}
    \field{year}{1948}
  \endentry

  \entry{Weizsacker48}{article}{}
    \name{author}{1}{}{%
      {{hash=vC}{%
         family={{von Weizs{\"a}cker}},
         familyi={v\bibinitperiod},
         given={{C.F.}},
         giveni={C\bibinitperiod},
      }}%
    }
    \list{publisher}{1}{%
      {Springer}%
    }
    \strng{namehash}{vC1}
    \strng{fullhash}{vC1}
    \field{labelnamesource}{author}
    \field{labeltitlesource}{title}
    \field{sortinit}{V}
    \field{sortinithash}{V}
    \verb{doi}
    \verb 10.1007/BF01668898
    \endverb
    \field{number}{7}
    \field{pages}{614\bibrangedash 627}
    \field{title}{Das Spektrum der Turbulenz bei großen Reynoldsschen Zahlen}
    \field{volume}{124}
    \field{journaltitle}{Zeitschrift f{\"u}r Physik}
    \field{year}{1948}
  \endentry

  \entry{Weizsacker51}{article}{}
    \name{author}{1}{}{%
      {{hash=vC}{%
         family={{von Weizs{\"a}cker}},
         familyi={v\bibinitperiod},
         given={{C.F.}},
         giveni={C\bibinitperiod},
      }}%
    }
    \strng{namehash}{vC1}
    \strng{fullhash}{vC1}
    \field{labelnamesource}{author}
    \field{labeltitlesource}{title}
    \field{sortinit}{V}
    \field{sortinithash}{V}
    \verb{doi}
    \verb 10.1086/145462
    \endverb
    \field{pages}{165}
    \field{title}{The Evolution of Galaxies and Stars}
    \field{volume}{114}
    \field{journaltitle}{The Astrophysical Journal}
    \field{year}{1951}
  \endentry

  \entry{Wetterich93}{article}{}
    \name{author}{1}{}{%
      {{hash=WC}{%
         family={Wetterich},
         familyi={W\bibinitperiod},
         given={C.},
         giveni={C\bibinitperiod},
      }}%
    }
    \strng{namehash}{WC1}
    \strng{fullhash}{WC1}
    \field{labelnamesource}{author}
    \field{labeltitlesource}{title}
    \field{sortinit}{W}
    \field{sortinithash}{W}
    \verb{doi}
    \verb 10.1007/BF01474340
    \endverb
    \field{pages}{451}
    \field{title}{The average action for scalar fields near phase transitions}
    \field{volume}{57}
    \field{journaltitle}{Z. Phys. C}
    \field{year}{1993}
  \endentry
\enddatalist

  \blx@bblend
  \endgroup
  \csnumgdef{blx@labelnumber@\the\c@refsection}{0}%
  \iftoggle{blx@reencode}{\blx@reencode}{}}

\newif\ifhyper
\hypertrue
\ifhyper
\hypersetup{
  citecolor = {green},
  colorlinks = {true}, 
  urlcolor = {blue} 
}
\fi

\newlength{\ldag}
	\bibliography{EcGSNSI.bib}

\begin{document}





\articletype{Paper}

\title{Energy Cascades in Driven Granular Liquids~: A new Universality Class? I~: Model and Symmetries}
\author{O. Coquand$^1$}

\affil{$^1$Laboratoire de Modélisation Pluridisciplinaire et Simulations, Université de
Perpignan Via Domitia, 52 avenue Paul Alduy, F-66860 Perpignan, France}

\email{oliver.coquand@univ-perp.fr}

\begin{abstract}
	This article deals with the existence and scaling of an energy cascade in steady granular liquid flows between the scale at which the
	system is forced and the scale at which it dissipates energy.
	In particular, we examine the possible origins of a breaking of the Kolmogorov Universality class that applies to Newtonian liquids
	under similar conditions.
	In order to answer these questions, we build a generic field theory of granular liquid flows and, through a study of its symmetries,
	show that indeed the Kolmogorov scaling can be broken, although most of the symmetries of the Newtonian flows are preserved.
\end{abstract}

%

%
%



\section{Introduction}

	The concept of \textit{granular matter} encompasses all the materials composed of non-Brownian particles interacting via dissipative collisions.
	Part of these systems are in a so-called \textit{fluid} state because they can flow --- the word "state" is used instead of "phase" here because, since
	the interactions between particles dissipate energy, granular matter can never be in thermodynamical equilibrium, except in the trivial
	state.
	Among these, the densest fluids are of particular interest.
	Indeed, due to gravity, most of granular flows on Earth, like debris flows, sand flows, landslides, mud flows for example, have a relatively high
	packing fraction (ratio of the total volume of the system occupied by the granular particles).

	Granular fluids with packing fractions in the range $0.4\lesssim\varphi\lesssim0.6$ are in the so-called \textit{liquid state} \cite{Andreotti13}.
	This state is separated from the more dilute \textit{gas state} by the presence of strong collective phenomena \cite{Coquand24},
	and from the \textit{solid state} by its ability to flow.
	In addition, very close to the solid state, there exist a \textit{friction dominated state} \cite{DeGiuli15,DeGiuli16,DeGiuli17a}, dominated by very
	slow \textit{creep flows}, which has a physics distinct both from the liquid and the solid, and governs, for example the physics of the \textit{jamming
	transition}.

	A point of particular interest in the physics of granular liquids is the existence of strongly universal laws governing the behavior of the flow.
	This enabled the discovery of a phenomenological law governing the rheology of granular liquids \cite{GDR04,DaCruz05} which is so robust that it has not
	really been challenged until now, more than twenty years after its discovery.
	There exists now a large corpus of experimental, numerical (see \cite{Forterre18} and references therein), as well as theoretical \cite{Kranz20,Coquand20f}
	studies confirming the ubiquity of this simple phenomenology.

	However, apart from their rheology, most of the properties of granular liquids are still elusive.
	This is mostly due to the fact that, as explained above, due to the dissipative character of their interactions, granular system always evolve
	far from equilibrium, so that most tools of statistical physics cannot be directly applied to study their properties.
	In particular, the way the energy is transported from the scales at which the system is forced to move to the scales at which it is dissipated, and its
	relations to the formation of mesoscopic structures within the granular liquid, which have been observed both numerically \cite{Radjai02,Saitoh16a}
	and experimentally \cite{Richefeu12,Combe15} remains unclear, to the best of our knowledge, today.
	This is the question we investigate in the present article.
	More precisely, we want to explore the possibilities of construction of a theoretical framework that could enable to tackle such questions
	from fundamental principles.

	In this study, we focus on the problem of \textit{simple shear flows}, assuming in addition that the system has reached a steady flow regime.
	The problem thus falls into the category of \textit{out of equilibrium steady states} for which a number of theoretical tools have been developed.
	In that case, the scale of energy injection is the scale at which the system is sheared, which corresponds to the largest macroscopic scale of the system,
	while the scale of energy dissipation, if the granular particles are not embedded into a viscous liquid, is of the order of a few particle
	diameters at best, and thus corresponds to the microscopic scale of the system.
	The question remains open to understand what occurs at intermediate scales, and how the injected energy flows into the system until it is finally dissipated.

	The problem of interaction between internal structures at different scales that carry energy in a continuous medium from a forcing scale into a dissipation
	scale is reminiscent of another well studied problem in fluid mechanics, namely that of the scaling in the so-called \textit{energy cascade} in
	homogeneous and isotropic fully developed turbulence of Newtonian liquids.
	In such systems, it was first shown by Kolmogorov \cite{Kolmogorov41a,Kolmogorov41b,Kolmogorov41c} that in an \textit{inertial range} going from the
	macroscopic forcing scale to a microscopic scale fixed by the viscosity of the fluid, the spectrum of the velocity fluctuations obeys a universal scaling
	law.
	This result can be interpreted in terms of the eddies that form in the turbulent flows as the existence of a \textit{cascade} of energy,
	that is a series of interactions between eddies of different sizes, whereby the larger scale eddies tend to give energy to the smaller
	energy ones, until the energy is finally dissipated at the microscopic scale.
	Interestingly, this property, particularly visible in flows of high Reynolds numbers (thus small viscosity), holds even in the limit where the
	viscosity of the fluid goes to zero.
	Indeed, it was already noted by Prandtl in 1926 that the interaction between eddies at the smaller scales should be interpreted not as a dissipation
	of the total energy by a form of effective friction between the smallest fluid cells, but rather as coming from the exchange of momentum between
	eddies of different sizes and velocities \cite{Prandtl26c}.

	Our strategy will thus be to use and adapt the very large corpus of studies of the scaling of energy cascades in turbulent flows in Newtonian
	liquids \cite{Prandtl26a,Prandtl26b,Weizsacker48,Heisenberg48,Onsager49,Weizsacker51,Batchelor58,Batchelor59,Kraichnan67,Kraichnan71,Kraichnan74a,
	Kraichnan74b,Kraichnan75,Falkovich94,Chertov95,Balkovsky99,Falkovich01,Eyink06,Falkovich11,Canet15,Canet16,Canet17,Tarpin19,Gorbunova20,Gorbunova21a,Canet22}
	to build up a framework adapted to the granular problem.
	In particular, we want to investigate how far the analogy between Newtonian and granular liquid flows goes, whether the definition of an energy cascade in the
	latter system is a relevant way to describe the physics at play, and if scaling laws exist, how different they are from the scaling predicted by the
	model of Kolmogorov (K41 scaling in the following).
	Indeed, granular liquids still retain a lot of properties from their Newtonian counterparts, and the prediction \cite{Coquand20g,Coquand21,Coquand23,Coquand24},
	and the observation \cite{Radjai02,Richefeu12,Combe15,Saitoh16a,Kharel17} of collective eddy-like structures in granular flows further enhances the power of the
	comparison.

	Before going further, let us summarize the main properties of fully developed homogeneous isotropic turbulence in Newtonian liquids~: (i) the fluid cell
	trajectories are chaotic \cite{Falkovich01}, thus the description of the velocity field, advected tensors and related quantities requires a statistical treatment;
	Physical observables are generally build from correlation functions of such fluctuating fields, (ii) in the inertial range, the K41 scaling governs the behavior of
	the correlation functions, (iii) a closer look at the behavior of the system, now possible thanks to the existence of better theoretical, numerical and experimental
	tools, reveals small deviations from K41, a phenomenon resulting from the existence of \textit{intermittency} in the system,
	(iv) there are strong effects of dimensionality; Indeed, in two dimensions,
	the \textit{enstrophy}, integral of the vorticity over the whole fluid, is a conserved quantity, so that the cascade of energy is reversed, and there exists
	a direct cascade of enstrophy that also participates to the momentum redistribution within the fluid, causing new scaling relations with, in particular,
	logarithmic terms \cite{Kraichnan67},
	(v) the properties of the advected quantities strongly depend on the smoothness properties of the velocity field, which themselves are related
	to the value of the fluid's viscosity (which provides an effective cutoff length scale below which the velocity field gets smoothened) \cite{Falkovich01},
	(vi) scaling laws outside of the inertial range may not be universal \cite{Balkovsky99}.

	As far as granular flows are concerned, on the other hand, the results are, to the best of our knowledge, very scarce, and still inconclusive due to the dispersion
	of results across the different studies.
	The first study has been conducted by Radjai and Roux \cite{Radjai02} by numerical simulation of a two dimensional shear flow,
	followed some years later by an experimental study of a two dimensional shear cell by Richefeu, Combe and their collaborators
	\cite{Richefeu12,Combe15} and finally, more numerical results have been obtained in the group of Saitoh \cite{Saitoh16a,Saitoh16,Oyama19,Saitoh20,Singh23}.
	Let us also mention some more numerical studies centered around the behavior of the diffusion coefficient across scales \cite{Kharel17,Artoni21}.
	Their results are as follows~: (i) Vortex-like structures similar to those observed in the Newtonian turbulence do exist in granular liquid flows,
	(ii) an energy cascade with anomalous scaling is present, but the value of the exponent is still unclear; In 3 dimensions it is reported to be -3/2,
	so different from the K41 prediction, but so far just one group has observed this value \cite{Oyama19}; In 2 dimensions, various values are reported.
	Radjai and Roux find a value of $-1.24$ identified with the fraction $-5/4$ while the group of Saitoh reports two values depending on the method used
	to evaluate the exponent~: $-6/5$ by a direct adjustment of the numerical data, and $-7/5$ by a finite size scaling analysis. While these exponents are
	different from that of Radjai and Roux, their exponent lies in the range defined by the two values given by Saitoh.
	Experimentally, the situation is less clear and one of the two studies \cite{Richefeu12} even reports a $-5/3$ scaling,
	in agreement with K41. It is to be noted, however, that part
	of their results have been obtained in the friction dominated regime, and not in the liquid regime that we propose to study.
	Given the complexity of the analysis of the turbulence spectrum in two dimensions already in the Newtonian case, we propose here to restrict ourselves to flows in
	three dimensions.
	Besides, contrary to the Newtonian case where two dimensional flows are of great importance for the study of climate phenomena for example, there is no
	analogous situation for granular liquids for which most experimentally relevant cases concern three dimensional flows; (iii) the fluctuations of the
	velocity field are strongly non Gaussian, and even more so that the packing fraction is increased.

	Our objective is to build a theoretical framework to investigate the relevance of the concept of energy cascades in granular liquids and more precisely
	if and how K41 scaling can be broken.
	Because we need a good description of scaling relations and how phenomena at different scales interact together, a rather natural language to express our
	formalism is that of field theory and the renormalisation group.
	Obviously, this approach has already been applied to Newtonian turbulence, and a lot of different approximation schemes for the renormalisation of the
	turbulence theory in the Navier-Stokes equation have been proposed (see \cite{Forster77,Falkovich94,Fedorenko13,Falkovich01} for examples).
	Here, we propose to adapt the method developed in the group of L. Canet that proved particularly powerful to tackle, not only the questions of K41
	scaling, but also the role of intermittency and scaling outside of the inertial range as well (see \cite{Canet22} for a review).
	On the granular side, we are going to use the Granular Integration Through Transients (GITT) approach, since it is, to the best of our knowledge,
	the only theory that allows to derive the behavior of the rheological observables in granular liquids.
	In particular, in \cite{Coquand21}, a general form for the viscosity tensor of a granular liquid, holding in all the different possible flow regimes,
	has been derived.
	The combination of a constitutive equation from GITT with a field theory describing the incompressible Navier-Stokes equation then yields a proper theory
	of granular liquid flows.
	Note that the coupling of the Navier-Stokes equation with a constitutive equation is the standard method used in simulation of geophysics flows
	on realistic topographies \cite{Kelfoun09}.

	The outline of the paper is as follows~: In a first section, we revisit qualitative arguments of von Weizsäcker and Heisenberg to get a first idea
	of why K41 can be broken in granular flows.
	In particular, we show that, using the robust numerical data about the scaling analysis of the diffusion coefficient of Kharel and Rognon, the
	-3/2 exponent observed by the group of Saitoh is recovered.
	Then, we explain how to build a field theory adapted to the description of our problem, combining tools from the description of scaling in
	dynamical systems and GITT.
	The constraints imposed by the form of the viscosity tensor on the possible expressions of the stress tensor, that explicitly appears in the Navier-Stokes
	equation, are studied in details.
	Finally, an extensive study of the symmetries of the effective action and the related Ward identities is conducted, as it has been shown that
	they give constraints sufficient to explain the existence of K41 in the inertial range \cite{Canet15}.
	Given that granular liquid flows inherit a lot of their properties from the Navier-Stokes equation, that is only slightly modified in their case,
	examining which symmetries are preserved and which ones are broken proves to be a very instructive exercise.
	The whole study is conducted in a most model-independent way so as to provide a frame in which different types of approximations for the study
	of the renormalisation group flow can be derived.
	The explicit design of such approximations, that inevitably have implications on the quality of the results, is kept for following articles.

\section{A Qualitative analysis of the energy cascade}

	Before diving into complex constructions, let us have a first estimate of the scaling laws in the energy cascade.
	These arguments are not fully rigorous, and in the granular case, we need to use some input from numerical data.
	Nonetheless, this first attempt shows that a breaking of K41 can be expected in granular liquids, notably due to the difference of nature
	of the dissipation mechanism, even though this one is not visible from the macroscopic scale at which the observer performs a measurement.

	\subsection{The argument of Von Weizsäcker}

		First, we follow the reasoning of von Weizsäcker in \cite{Weizsacker48}.
		After reminding the argument in the Newtonian case, we derive its granular equivalent.

		\subsubsection{The argument for Newtonian liquids}

			The following section reproduces briefly the argument of \cite{Weizsacker48}. The interested reader is referred to the original paper for
			more details.

			The derivation of the -5/3 law for the energy spectrum in this reasoning lies upon two crucial points~: (i) The scaling relations
			need to be derived taking into account interactions between structures at various length scales in the fluid; (ii) The Prandtl
			Misschungsweghypothese holds, namely, there exists a length scale equivalent to the mean free path in kinetic gas theory,
			but defined for two elementary cells
			of liquid \cite{Prandtl26c}.

			Let us define a series of length scales $\big\{L_i\big\}_{i\in\mathbb{N}}$ such that,
			\begin{equation}
				\frac{L_{n+1}}{L_n}= \delta\,,
			\end{equation}
			with $0<\delta<1$ being a constant coefficient that needs not to be specified.
			From this hierarchy of lengths, one can define the root mean squared velocity at length $L_n$, $v_n$, and the root mean squared
			velocity gradient at length $L_n$, $v'_n$ (the means here are time averages).
			Then, if $L_n$ is sufficiently large compared to the smallest eddies size and sufficiently small compared to the largest eddies size,
			it is expected that
			\begin{equation}
				v_n'=\alpha\,\frac{v_n}{L_n}\,.
			\end{equation}
			If, moreover, the medium is assumed to be homogeneous, it is expected that $\alpha$ does not depend on $n$.

			The resulting viscosity at scale $n$ can then be expressed as follows~:
			\begin{equation}
			\label{eqPreta}
				\eta_n = \rho\,l_n\,w_n\,,
			\end{equation}
			where $\rho$ is the liquid's density (homogeneous, so supposed to be scale independent),
			$l_n$ is Prandtl's mixing length (Misschungsweg in German) and $\displaystyle w_n = \sum_{i=n+1}^{+\infty}v_i$.
			Following the same line of reasoning as above, it is expected that
			\begin{equation}
				l_n=\beta\,L_n\,,
			\end{equation}
			where $\beta$ does not depend on $n$ if the medium is homogeneous.

			The dissipated energy at scale $n$, denoted $S_n$, is given by the viscosity times the average of the square of the rotational of the velocity
			field over time~:
			\begin{equation}
			\label{eqSn}
				S_n=\eta_n\,\gamma\,\sum_{i=0}^n \big(v'_i\big)^2
			\end{equation}
			where again, $\gamma$ is some constant independent of $n$.
			The fact that only the smallest values of $k$ enter the sum directly stems from the definition of the energy cascade: the energy
			flows from the (low $k$) large scales structures down to the smaller scales (larger $k$) structures, until the Kolmogorov scale is reached.

			Searching for an ansatz of the form
			\begin{equation}
				v_{n+1} = \zeta\,v_n\,,
			\end{equation}
			it is not difficult to prove that
			\begin{equation}
				w_{n+1}= \frac{\zeta}{1-\zeta}v_n\ ,\ \sum_{i=0}^n\big(v'_i\big)^2= \frac{\alpha^2}{1-(\delta/\zeta)^2}\left(\frac{v_n}{L_n}\right)^2
			\end{equation}

			Now, if the energy cascade obeys a well-defined power law scaling, then $S_n$ must be independent of $n$, what translates into
			\begin{equation}
				S_n\propto\eta_n\left(\frac{v_n}{L_n}\right)^2\propto\frac{v_n^3}{L_n}
			\end{equation}
			independent of $n$, namely,
			\begin{equation}
				v_n\propto L_n^{1/3}
			\end{equation}
			Finally, the kinetic energy $\mathcal{E}_n$ scales as
			\begin{equation}
				\mathcal{E}_n\propto v_n^2\propto L_n^{2/3}\,,
			\end{equation}
			so that its Fourier transforms becomes
			\begin{equation}
				\mathcal{E}_k=\int_k^{+\infty}F(k')dk'\propto k^{-2/3}\ \Rightarrow\ F(k)\propto k^{-5/3}
			\end{equation}
			which is nothing but the K41 law for the scaling of the spectrum of the kinetic energy.

		\subsubsection{Adaptation to the granular case}

			Among the laws derived above, some are general truths, others have no reason to change when the nature of the fluid changes.
			In fact, the only difference between Newtonian and granular liquids from this point of view is the mechanism of dissipation of energy.
			For simplicity, we restrict ourselves to the case where collisions are the only source of energy dissipation. In that case,
			the dissipated energy at scale $n$ can be written as
			\begin{equation}
				S_n^G = \rho\,\Gamma_d\,\omega_n\,T_n\,,
			\end{equation}
			where $\Gamma_d$ is a constant damping rate depending only on the restitution coefficient in the collisions (a microscopic quantity
			fully independent on scale), $\omega_n$ is the extension of the
			collision frequency at scale $n$, and $T_n$ is the granular temperature at scale $n$.
			Pay attention to the fact that, in granular systems, the temperature is purely kinetic and not a thermodynamic one, it is defined formally
			from the second moment of the velocity fluctuation correlation function.

			The concept of "collision frequency at scale $n$" can be deemed a little bit peculiar.
			In fact, it should rather be interpreted as a collision frequency between structures at scale $n$ built in such a way that
			when $n$ is such that $L_n$ equals to one particle diameter, $\omega_n$ is equal to the usual collision frequency between two particles.

			In order to go further, we need two more things.
			First, we need to define $\dot\gamma_n$, the effective strain rate at scale $n$.
			It can be related to the total strain rate $\dot\gamma_0$, defined at the macroscopic scale and present in the rheological laws.
			In order to do so, we use the following relation, observed numerically \cite{Singh23}~:
			\begin{equation}
				\left<V_y\right>_n \propto l_n \,\dot\gamma_n\,,
			\end{equation}
			where $V_y$ is the total velocity, including the average flow, $y$ is the direction of shear, and $l_n$ is a typical size of the
			vortex-like structures that develop in the granular flow (for the same reasons as above, $l_n\propto L_n$).

			It is not difficult to establish that~:
			\begin{equation}
				\left<V_y\right> = \frac{1}{L_0}\int_0^{l_n} y \dot\gamma_0\,dy = \frac{l_n^2}{2L_0}\dot\gamma_0\,.
			\end{equation}
			Thus, we deduce the relationship between the macroscopic and mesoscopic strain rates~:
			\begin{equation}
				\dot\gamma_n\propto L_n\,\dot\gamma_0
			\end{equation}
			This is the granular equivalent of the Prandtl Misschungsweghypothese (that does not hold anymore in that case).
			Here, we replaced the concept of mixing length by that of an effective, scale dependent, strain rate, that describes the strength of strain
			at the scale of the structure under study.
			Note that in the above equation, $\dot\gamma_0$ is fixed by the environment, and is therefore a constant, independent of $n$.

			Second, we need to relate $\omega_n$ to the velocity fluctuations.
			In the following, it should be understood that $v_n$ is the velocity fluctuation field, from which the average velocity field imposed by the
			simple shear flow has been subtracted, so that its direction does not need to be specified.
			There is, to the best of our knowledge, no obvious scaling relation that could allow us to get such a relation.
			Therefore, we chose to use the solidly established scaling law observed numerically in \cite{Kharel17}~:
			\begin{equation}
			\label{eqRog}
				\frac{v_n}{\dot\gamma_n}\propto \mathcal{I}_n^{-1/2}\propto\left(\frac{\omega_n}{\dot\gamma_n}\right)^{1/2}
			\end{equation}
			where $\mathcal{I}$ is the inertial number.
			It has been proven in a previous study that the inertial number can be expressed, up to an irrelevant constant, as the ratio
			of the shear rate by the collision frequency $\omega_n$ \cite{Coquand20g}, which justifies the second equality in (\ref{eqRog}).
			From this, we deduce
			\begin{equation}
				\omega_n\propto\frac{v_n^2}{\dot\gamma_n}
			\end{equation}
			Finally, remembering that, by definition, $T_n\propto v_n^2$,
			\begin{equation}
				S_n^G\propto\frac{v_n^4}{L_n}\ \Rightarrow\ v_n\propto L_n^{1/4}
			\end{equation}
			where for the last equality, we used the hypothesis that the dissipation rate should be independent of the scale.

			As for the kinetic energy spectrum, we therefore get
			\begin{equation}
				\mathcal{E}_k^G=\int_k^{+\infty}F(k')dk'\propto k^{-1/2}\ \Rightarrow\ F(k)\propto k^{-3/2}
			\end{equation}
			Interestingly, this is precisely the value reported by the group of Saitoh in the only 3 dimensional study of the energy spectrum \cite{Oyama19}.
			Let us stress that the scaling relation (\ref{eqRog}) we used in the derivation was obtained completely independently in a different group.

			All in all, even though granular liquids share a lot of properties with Newtonian fluids, in the regime where collisions dominate the
			dissipation, a new scaling systems emerges that differs from K41.
			Let us emphasize that we do not pretend having proved anything so far, but rather hinted at a serious reason why a different set of scaling
			laws can be expected in granular liquids.

			Lastly, let us discuss another qualitative argument binding results across different approaches that hints towards the -3/2 scaling as well.
			In \cite{Kranz14}, a study of the behavior if the velocity autocorrelation function $\psi(t)$ at large times has been conducted by the
			use of a mode coupling approximation model.
			The velocity autocorrelation function is the two point time correlation of the velocity of a \textit{tagged particle} in the medium,
			thus it has no spatial argument, which greatly simplifies the writing of the mode coupling dynamical equation.
			In that model, it was shown that \cite{Kranz14}~:
			\begin{equation}
				\psi(t)\underset{t\rightarrow+\infty}{\sim}t^{-3/2}
			\end{equation}
			Now, let us examine this statement in the light of the numerical study \cite{Kharel17}, chosen for the quality of the presented data.
			In this article, it has been observed that the spatial dispersion at scale $n$, $\left<\Delta x_i^2\right>_n$ --- because we always study
			fluctuating quantities from which the background flow has been subtracted, the direction $i$ does not matter --- follows a modified
			diffusion equation
			\begin{equation}
				\left<\Delta x_i^2\right>_n = D_n\,t_n
			\end{equation}
			where $D_n$ is a scale dependent diffusion coefficient, and $t_n=1/\dot\gamma_n\propto 1/L_n$
			(be aware that the notations used here may differ a bit from the original work \cite{Kharel17} for need of homogenisation of the notations).
			Moreover,
			\begin{equation}
				D_n \propto \frac{1}{t_{v,n}}\,,
			\end{equation}
			where $t_{v,n}$ is a time scale associated with the dynamics of the vortices at scale $n$.
			It is further shown numerically in the same work that
			\begin{equation}
				t_{v,n}\propto\frac{1}{\dot\gamma_n\,L_n}\propto L_n^{-2}\,.
			\end{equation}
			Thus, we can make the rough estimate
			\begin{equation}
				\psi_n(t_n)\sim\dot\gamma_n^2\left<\Delta x_i\right>_n^2\propto L_n^3\propto \big(t_{v,n}\big)^{-3/2}
			\end{equation}
			Hence, we can expect from \cite{Kharel17}, shown to be a priori consistent with the -3/2 decay of the energy spectrum by the argument above,
			that the scale dependent velocity autocorrelation function decays as the -3/2 power of a time scale, which is nothing but the characteristic
			time scale of the vortex dynamics.
			It is interesting, in our opinion that this prediction is fully consistent with the study based on the mode coupling approximation model, on a
			completely independent basis, which reinforces the expectation that the collision dynamics is the key to the deviation away from the K41
			scaling in granular liquids.

			Finally, it is not uninteresting to note that the 3/2 law has been predicted in a different context: In magnetohydrodynamic turbulence,
			it corresponds to the so-called Iroshnikov-Kraichnan law \cite{Iroshnikov64,Kraichnan65}.
			In \cite{Iroshnikov64}, it is shown that when considering collision between two magnetohydrodynamic waves in the liquid, with an arbitrary
			interaction kernel, the energy spectrum cannot be independent of scale unless the power spectrum scales as $k^{-3/2}$.
			Of course, the physics of interaction between magnetic and viscous effects in a liquid cannot be easily mapped onto the problem of the
			sheared granular liquid; However, the fact that the power spectrum seems to scale with the exact same exponent raises the question of
			the possible links that can be drawn between colliding magnetohydrodynamic waves in a liquid and the structures having an effective
			collision frequency of $\omega_n$ in the granular liquid. This subject surely deserves further investigation.

	\subsection{The refinement of Heisenberg}

		In a subsequent paper \cite{Heisenberg48}, Heisenberg proposed a refined analysis of the scaling relations that allowed to predict a scaling in the
		higher $k$ regime, out of the inertial range.
		Even though the predicted scaling has not since been solidly established \cite{Betchov57}, we still propose to analyse the argument to see can be learnt from it
		regarding the granular case.

		\subsubsection{The Case of Newtonian Liquids}

			In his demonstration, Heisenberg uses a continuous representation in Fourier space for the scale dependence, rather than the discrete
			version of von Weizsäcker.
			The principle is the same, the notations we use are the natural extensions of those used in the above section.

			The energy spectrum function $F$ can be expressed in terms of the Fourier decomposition of the velocity fluctuation modes as~:
			\begin{equation}
				F(k) = \frac{V}{(2\pi)^2}\,k^2\,v_k^2
			\end{equation}
			where $V$ is the total volume of the system.
			The dissipated energy (that, as shown above does not depend on the scale if a proper power law scaling regime is established) then becomes~:
			\begin{equation}
			\label{eqSHei}
				S = \big(\eta_B+\eta_k)\int_0^k\,2k'^2F(k')dk'
			\end{equation}
			where $\eta_B$ is a bare viscosity, introduced by Heisenberg, so that $\eta_k$ only represents viscosity fluctuations at scale $k$.
			Thus, by definition, the bare viscosity $\eta_B$ corresponds to the value of the viscosity in the limit of large wave numbers, 
			which should saturate at a finite value.
			The integral from 0 to $k$ is to be compared to the sum over large scale modes in Eq.~(\ref{eqSn}) is von Weizsäcker's model.

			Now, according to Prandtl's Misschungsweghypothese, $\eta_k$ can be expressed as
			\begin{equation}
				\eta_k\propto L_k\,v_k\propto L_k^{4/3}
			\end{equation}
			or,
			\begin{equation}
				\eta_k = \kappa\rho\int_k^{+\infty}dk'\left(\frac{F(k')}{k'^3}\right)^{1/2}\,,
			\end{equation}
			to be compared with Eq.~(\ref{eqPreta}), where a sum over the small length scale modes is involved.
			Here, $\kappa$ is a mere constant prefactor that does not play any special role.

			Writing that
			\begin{equation}
				\frac{dS}{dk}=0\,,
			\end{equation}
			using the notations
			\begin{equation}
				\eta_B=\nu\rho\ ,\ x=\ln\left(\frac{k}{k_0}\right)\ ,\ F(k)=F(k_0)e^{-w(x)}\,,
			\end{equation}
			and identifying the Reynolds number $\text{Re}$ in
			\begin{equation}
				\frac{\nu\sqrt{k_0}}{\kappa\sqrt{F(k_0)}} = \frac{\text{Re}_0}{\text{Re}}\,,
			\end{equation}
			one is finally lead to
			\begin{equation}
				\begin{split}
					&\left(\frac{7}{2}-\frac{1}{2}\frac{dw}{dx}\right)\left(\frac{\text{Re}_0}{\text{Re}} + \int_x^{+\infty}e^{-w(x')/2-x'/2}dx'\right) \\
					&= 2\,e^{-w(x)/2-x-2}
				\end{split}
			\end{equation}

			Since the reasoning is based on the hypothesis that a proper scaling regime exists, it is expected that ${w(x)\simeq Wx}$, so that the
			integral can be approximated by its steepest descent value, which finally gives~:
			\begin{equation}
				\big(7-W\big)\left(\frac{\text{Re}_0}{\text{Re}}+e^{-(W-1)x/2}\frac{2}{1+W}\right)=4\,e^{-(W-1)x/2}
			\end{equation}

			In the very high Reynolds regime, the first term in the parenthesis can be neglected, so that
			\begin{equation}
				7-W = 2\big(1+W\big)\ \Rightarrow\ W=\frac{5}{3}\,.
			\end{equation}
			This corresponds to the K41 regime.

			If, to the contrary, $k\gg k_0$, $x\gg1$, so that the equality can only hold if
			\begin{equation}
				W\simeq 7
			\end{equation}

			Hence, the derivation of Heisenberg allows to get an estimate of the power law scaling at short distances, outside of the inertial range.

		\subsubsection{The Case of Granular Liquids}

			Now, let us adapt the above calculation to the case of granular liquids.
			As we explained before, the only notable difference lies in the energy dissipation equation.
			Once again, we will restrict ourselves to the case where dissipation by collisions dominate.

			Following our preceding derivation, the analogous equation of Eq.~(\ref{eqSHei}) in the granular case is
			\begin{equation}
				S^G=\big(\eta_B+\eta_k\big)\,T_k
			\end{equation}
			with
			\begin{equation}
				T_k=\int_0^k F(k')\,k'^{\theta}dk'\sim k^{\gamma}\,.
			\end{equation}
			We also know that
			\begin{equation}
				T_k\propto v^2_k\propto k^{-1/2}\,,
			\end{equation}
			so that $\gamma = -1/2$ .

			For the viscosity, we do not know a priori the expression of the integral over the small length scales, so we use
			generic exponents $\alpha$ and $\beta$~:
			\begin{equation}
				\eta_k=\kappa\rho \int_k^{+\infty} k'^\alpha F(k')^\beta dk'\,.
			\end{equation}
			Using the fact that $\eta_k\propto k^2$ in that case, we get the relation~:
			\begin{equation}
			\label{eqalbe}
				\alpha = 1+\frac{3\beta}{2}\,.
			\end{equation}
			We can now proceed along the same steps, using that
			\begin{equation}
				\frac{dS^G}{dk}=0\,,
			\end{equation}
			which yields
			\begin{equation}
				\begin{split}
					&\left(\frac{\text{Re}_0}{\text{Re}}+e^{(-\beta W + \alpha+1)x}\frac{1}{W-(\alpha+1)}\right) \\
					&\times\left((\beta-1)W-\left(\alpha+\frac{1}{2}
							\right)\right)\simeq 2e^{(-\beta W + \alpha+1)x}
				\end{split}
			\end{equation}

			In the high Reynolds regime, the same reasoning as above gives
			\begin{equation}
				W=\frac{5+3\beta}{2(3-\beta)}\,,
			\end{equation}
			which is compatible with the von Weizsäcker's result $W=3/2$ if
			\begin{equation}
				\beta=\frac{2}{3}\ ,\ \alpha = 2\,.
			\end{equation}

			On the other hand, in the small length scales regime,
			\begin{equation}
				(\beta-1)W-\left(\alpha + \frac{1}{2}\right)=0\ \Rightarrow\ W=-\frac{15}{2}\leqslant0
			\end{equation}
			Clearly, the negative value for $W$ goes against our hypotheses because it then means that the Reynolds term
			cannot dominate in the $x\gg1$ regime.
			We thus come to the conclusion that --- up to the fact that our reasoning so far is only qualitative and not rigorous --- there can be
			no power law scaling regime in the small length scale region outside of the inertial range for granular liquids.
			This result is very interesting if one thinks about the main difference between granular and Newtonian liquids~: In the former case,
			the elementary particles are big (bigger than 100 $\mu$m typically), so that when going to smaller and smaller length scales, the continuous
			medium hypothesis breaks down sooner.
			Another way to phrase this is that in Newtonian liquids, the mechanism for energy dissipation is identified as a momentum transfer between
			the structures at the viscous scale, where the liquid is still considered as continuous, and not a molecular assembly; For granular liquids
			to the contrary, the scale of energy dissipation is of the order of magnitude of one particle diameter and the continuous hypothesis
			breaks down if we go to smaller length scales.
			The impossibility of existence of a power law scaling regime at short lengths can then be understood as a signature of the breakdown
			of the continuous medium hypothesis in the writing of the dissipation by collisions term within the continuous medium theory.
			Let us also remind that the above result has been derived under the hypothesis that dissipation by collisions dominates over all the other mechanisms
			of energy dissipation.

\section{A field theory for sheared granular matter}

	We now derive a theoretical framework that can be used to study the energy cascades in complex liquids, and particularly granular liquids.
	Our approach is built to be as model-independent as possible to be able to discuss the relation to the K41 scaling on the most general grounds,
	without depending too much on one particular set of approximations.

	\subsection{Dynamical equations for incompressible flows of granular liquids}

		The basis for the study of the dynamics of complex liquids is the momentum conservation energy, that closely looks like the Navier-Stokes equation,
		except that the shear stress $\sigma_{\alpha\beta}$ is a priori a generic \textit{functional} of the velocity fluctuation field $\mathbf{v}$~:
		\begin{equation}
		\label{eqSNS}
			\partial_t v_\alpha + v_\beta \partial_\beta v_\alpha = \partial_{\beta}\big(\sigma_{\alpha\beta}[\mathbf{v}]\big) + f_\alpha\,,
		\end{equation}
		where $f_\alpha$ is a random forcing term that encompasses all the energy injection processes needed for the out of equilibrium steady state to
		be maintained, and we chose a system of units where $\rho=1$ to lighten notations ($\rho$ is not renormalised and does not play any meaningful
		role in the following).
		The stochastic force is chosen to have a zero mean $\left<f_\alpha\right>=0$, and a variance profile that is local in time, and quite general in space~:
		$\left<f_\alpha(\mathbf{x},t)f_{\beta}(\mathbf{x}',t')\right> = 2\delta(t-t')N_{L,\alpha\beta}\big(|\mathbf{x}-\mathbf{x}'|\big)$.
		The index $L$ in the definition of the space profile is here to recall that it should be centered around a typical length scale $L$, that is
		usually a macroscopic one for the typical situations we describe.
		Generally, the study of the energy cascade requires a well-defined scale separation between $L$ and the microscopic scale of the system
		(for example, the diameter of a granular particle).
		The pressure term is part of the stress tensor and is not specified independently for now.
		As it stands, Eq.~(\ref{eqSNS}) is therefore quite similar to that of the well studied stochastic Navier-Stokes (SNS) model.

		In order to simplify our computations, we will further suppose that the fluid is incompressible, a very reasonable assumption for granular liquids
		once the steady flow is properly defined (and more generally for liquids that are by definition in a condensed phase), which leads to the further constraint~:
		\begin{equation}
		\label{eqICom}
			\partial_\alpha v_\alpha = 0\,.
		\end{equation}
		This means that the velocity field is purely transverse in the momentum space, a property that has profound implications on the possible choices of forms of the
		stochastic terms in the Navier-Stokes equation (see below).

	\subsection{The generators of the correlation functions}

		In order to be able to compute correlation functions from Eqs.~(\ref{eqSNS}) and (\ref{eqICom}), we first use the Martin-Siggia-Rose-Janssen-de Dominicis
		formalism \cite{Martin73,Janssen76,Dominicis76} (MSRJD for short in the following), which consists in defining the following generating functional
		(we use the notations of \cite{Canet16})~:
		\begin{equation}
		\label{eqZ}
			\begin{split}
				\mathcal{Z}\big[\mathbf{J},\overline{\mathbf{J}},K,\overline{K}\big]&= \!\!\!\int\mathcal{D}[\mathbf{v}]\mathcal{D}[\overline{\mathbf{v}}]
				\mathcal{D}[p]\mathcal{D}[\overline{p}]e^{-\mathcal{S}_\Lambda[\mathbf{v},\overline{\mathbf{v}},p,\overline{p}]
				-\Delta\mathcal{S}_\Lambda[\mathbf{v},\overline{\mathbf{v}}] } \\
												    &\times e^{+\mathbf{J}\cdot\mathbf{v}
													    +\overline{\mathbf{J}}\cdot
												    \overline{\mathbf{v}}+K\cdot p
											    +\overline{K}\cdot\overline{p}}\,,
			\end{split}
		\end{equation}
		where $\mathbf{J}$, $\overline{\mathbf{J}}$, $K$ and $\overline{K}$ are the sources, the pressure field $p$, defined in incompressible fluids
		by the relation $\displaystyle \sigma_{\alpha\beta} = -p\,\delta_{\alpha\beta} + \Delta\sigma_{\alpha\beta}$ ($\sigma$ being the stress tensor, and
		Tr$(\Delta\sigma)=0$.), has been used explicitly for reasons that will
		be explained later, $\overline{\mathbf{v}}$ and $\overline{p}$ are the velocity and pressure response fields respectively, the microscopic action
		$\mathcal{S}_\Lambda$ ($\Lambda \sim L^{-1}$) is given by
		\begin{equation}
		\label{eqSLambda}
			\begin{split}
				\mathcal{S}_\Lambda[\mathbf{v},&\overline{\mathbf{v}},p,\overline{p}] =\!\! \int_{\mathbf{x},t}\!\!\!\Big\{\overline{v}_{\alpha}(\mathbf{x},t)
					\big(\partial_tv_{\alpha}(\mathbf{x},t)+v_{\beta}(\mathbf{x},t)\partial_\beta v_\alpha(\mathbf{x},t) \\
							       & +\partial_\alpha p(\mathbf{x},t)-\partial_{\beta}\Delta\sigma_{\alpha\beta}[\mathbf{v}]
					       \big)
				       + \overline{p}(\mathbf{x},t)\partial_\alpha v_\alpha(\mathbf{x},t)\Big\} \,,
			\end{split}
		\end{equation}
		where $\Delta\sigma$ is the stress tensor from which the pressure of an equivalent \textit{quiescent} liquid at the same temperature (kinetic temperature in the
		granular case) has been subtracted, the notation $\int_{\mathbf{x},t}=\int d^3x\,dt$ has been introduced, the action correction from the stochastic
		terms is
		\begin{equation}
			\Delta\mathcal{S}_\Lambda[\mathbf{v},\overline{\mathbf{v}}]=-\int_{\mathbf{x},\mathbf{x}',t}\overline{v}_\alpha(\mathbf{x},t)
			N_{L,\alpha\beta}\big(|\mathbf{x}-\mathbf{x}'|\big)\overline{v}_\beta(\mathbf{x}',t)\,,
		\end{equation}
		and we used the shorthand notation ${U\cdot\phi=\int_{\mathbf{x,t}}U(\mathbf{x},t)\phi(\mathbf{x},t)}$ for the sources.
		In order to generate such a result, an average over the stochastic force has been performed.

		For the study of the interaction of physical quantities across scales, the generating functional has to be made scale-dependent, which
		allows then for a renormalisation group treatment.
		There are many ways to do so, for a review of the perturbative approaches and their limitations, the reader is referred to \cite{Canet16}.
		As explained above, we use here the effective average action formalism developed in the case of Newtonian turbulence in the group of L. Canet
		and N. Wschebor.
		It consists in replacing the action correction term by regulator term of the form~:
		\begin{equation}
			\begin{split}				
				\Delta \mathcal{S}_k[\mathbf{v},\overline{\mathbf{v}}]&=-\int_{\mathbf{x},\mathbf{x}',t}\overline{v}_\alpha(\mathbf{x},t)
				N_{k,\alpha\beta}\big(|\mathbf{x}-\mathbf{x}'|\big)\overline{v}_\beta(\mathbf{x}',t) \\
											      &+\int_{\mathbf{x},\mathbf{x}',t}\overline{v}_\alpha(\mathbf{x},t)
				R_{k,\alpha\beta}\big(|\mathbf{x}-\mathbf{x}'|\big)v_\beta(\mathbf{x}',t)
			\end{split}
		\end{equation}
		where the $k$-dependent terms are such that, if $q$ is the wavenumber corresponding to the Fourier representation of the cutoff function
		$R_k(q)$, in the limit $q\ll k$, where $q$ is the scale at which the system is examined, $R_k(q)\sim k^2$,
		so that the small wavenumber modes are damped, and in the other limit $q\gg k$, $R_k(q)\rightarrow 0$ and $N_k\rightarrow0$.
		Thus, $R_k$ and $N_k$ effectively act as selective cutoffs in the procedure of step by step fluctuation integration defined by the renormalisation
		group procedure~: When these terms are present, the high momentum fluctuation modes are integrated while the low momentum, large scales, 
		fluctuation modes are frozen.
		In that way, $\mathcal{Z}$ is upgraded to a scale-dependent quantity in which only the fluctuation modes of momentum superior to $k$ are
		taken into account.
		In the limit $k\rightarrow0$, this means that all fluctuation modes are integrated, namely, the physical quantities, measurable at the macroscopic
		scale, are recovered.

		In principle, the procedure does not depend on the specific choice of the functions $R_k$ and $N_k$, as long as the above restrictions are properly
		applied.
		In practice however, the quality of an estimate always depends on the quality of the approximation scheme that is designed to tackle the problem.
		In order to avoid generating artifacts, it is generally crucial that the cutoff functions $R_k$ and $N_k$ respect the fundamental symmetries of the
		system as much as possible.

		Symmetries can also be a source of simplification in the choice of these functions.
		Indeed, because the flow is incompressible, the velocity vector is purely transverse in momentum space.
		As a result, $N_{k,\alpha\beta}$ and $R_{k,\alpha\beta}$ can be chosen proportional to $\delta_{\alpha\beta}$, which greatly
		simplifies their expression \cite{Canet16}.
		Since we do not want to discuss any particular approximation scheme in this paper, we do not discuss further the properties of the cutoff functions.

		Replacing $\Delta \mathcal{S}_\Lambda$ by $\Delta \mathcal{S}_k$ in Eq.~(\ref{eqZ}) allows to define a scale dependent generating functional for the
		moments of the field $\mathcal{Z}_k$.
		However, this functional is not the most useful one.
		First, taking a logarithm allows to define a free energy at scale $k$, $\mathcal{W}_k$, which is the generating functional of the \textit{connected}
		correlation functions~:
		\begin{equation}
			\mathcal{W}_k\big[\mathbf{J},\overline{\mathbf{J}},K,\overline{K}\big] = -\,\ln\big(\mathcal{Z}_k\big[\mathbf{J},
			\overline{\mathbf{J}},K,\overline{K}\big]\big)\,,
		\end{equation}
		from which we can derive the macroscopic expectation values of the fields
		\begin{equation}
			\begin{split}
				& u_\alpha(\mathbf{x},t) = \left<v_\alpha(\mathbf{x},t)\right> = \frac{\delta W_k}{\delta J_\alpha(\mathbf{x},t)}\bigg|_{U=0} \\
				& P(\mathbf{x},t) = \left<p(\mathbf{x},t)\right> = \frac{\delta W_k}{\delta K(\mathbf{x},t)}\bigg|_{U=0}\,,
			\end{split}
		\end{equation}
		(where we recall that $U$ stands for a generic source), and of the response fields
		\begin{equation}
			\begin{split}
				& \overline{u}_\alpha(\mathbf{x},t) = \left<\overline{v}_\alpha(\mathbf{x},t)\right> = \frac{\delta W_k}{\delta
				\overline{J}_\alpha(\mathbf{x},t)}\bigg|_{U=0} \\
				& \overline{P}(\mathbf{x},t) = \left<\overline{p}(\mathbf{x},t)\right> = \frac{\delta W_k}{\delta \overline{K}(\mathbf{x},t)}\bigg|_{U=0}\,.
			\end{split}
		\end{equation}
		Then, we can perform a Legendre transform to compute the effective average action $\Gamma_k$, functional of the field expectation values~:
		\begin{equation}
			\begin{split}
				\Gamma_k\big[\mathbf{u},\overline{\mathbf{u}},P,\overline{P}\big] &= - \mathcal{W}_k\big[\mathbf{J},\overline{\mathbf{J}},K,\overline{K}\big]
				+ \mathbf{J}\cdot\mathbf{u}+ \overline{\mathbf{J}}\cdot\overline{\mathbf{u}}\\
												  & +K\cdot P + \overline{K}\cdot\overline{P}
												  -\overline{\mathbf{u}}\cdot R_k\cdot\mathbf{u}
												  + \overline{\mathbf{u}}\cdot N_k\cdot\overline{\mathbf{u}}
			\end{split}
		\end{equation}
		that is also the generating functional of the cumulants of the fluctuating quantities at scale $k$.
		In order to lighten notations, we will in the following condense $N_{k,\alpha\beta}$ and $R_{k,\alpha\beta}$ into only one tensor,
		denoted $N_k$ for short.

		Finally, the renormalisation group flow equation of $\Gamma_k$ is known exactly \cite{Wetterich93}~:
		\begin{equation}
		\label{eqWe}
			\partial_k\Gamma_k = \frac{1}{2}\text{Tr}\int_\mathbf{q}\partial_k\mathcal{R}_k(\mathbf{q})\cdot G_k(\mathbf{q})\,,
		\end{equation}
		where our asymmetric convention for the Fourier transform writes $\int_\mathbf{q}=\int d^nq/(2\pi)^n$, the renormalised propagator is
		\begin{equation}
			G_k(\mathbf{q}) = \Big[\Gamma_k^{(2)} + \mathcal{R}_k\Big]^{-1}\,,
		\end{equation}
		the superscript $(2)$ indicates a second order functional derivative with respect to the fields, and ${\mathcal{R}_k(\mathbf{q})=\Delta\mathcal{S}^{(2)}_k
		(\mathbf{q})}$.

		Obviously, Eq.~(\ref{eqWe}) as it stands is too general, and a proper approximation scheme has to be designed before this equation is solved.
		However, as far as our work in this article is concerned, we do not need to be more specific about how elements are kept or truncated.
		The main force of this formulation of the renormalisation group flow is that it opens the road to multiple ways of making approximations,
		both perturbative, and non-perturbative.
		In particular, in following papers we will show how a minimal model can be built, and how it can be refined in order to increase our degree of precision.
		Something that we still miss, though, is a proper expression for the stress tensor $\Delta\sigma$.
		This is the topic of the next section.
		Note that this means that all of the above can be applied to arbitrary models of (incompressible) complex liquid flows, independent on whether they
		are composed of granular particles or not (it can be applied as it stands for the study of flows of dense colloidal suspensions for example).

		\subsection{Accounting for the constitutive equation}

			In this section, we recall the main properties of the viscosity tensor in the GITT model.
			The GITT approximation is a model of granular liquids based on mode coupling theory \cite{Goetze08}, which accounts for the
			non Newtonian behavior of the liquid in the appropriate regimes of packing fraction and temperature, as well as
			the so-called integration through transients formalism \cite{Fuchs02,Fuchs03,Fuchs09} that enables to estimate statistical averages in the
			out of equilibrium sheared system.
			The derivation of the main equations is very long and tedious; Thus we preferred to discuss in this paper only the properties of the viscosity
			tensor.
			For more information about the building of the GITT equation and their resolution, the reader is referred to \cite{Kranz18,Kranz20,Coquand20f,
			Coquand20g}; For the derivation of the expression of the viscosity tensor from the GITT equations, see \cite{Coquand21};
			For the experimental validation of the GITT model, see \cite{Angelo23,Angelo25}.

			In the integration through transients formalism, statistical averages in the sheared system, denoted $\left<\cdot\right>^{(\dot\gamma)}$,
			are expressed in terms of quantities involving only averages in a fictitious, quiescent, unsheared liquid at same temperature and
			pressure, denoted $\left<\cdot\right>^{(0)}$.
			For the stress tensor, this leads to~:
			\begin{equation}
			\label{eqSigGITT}
				\left<\sigma_{\alpha\beta}\right>^{(\dot\gamma)} = \left<\sigma_{\alpha\beta}\right>^{(0)} + \Lambda_{\alpha\beta\theta\nu}(\dot\gamma)
				\kappa_{\theta\nu}\,,
			\end{equation}
			where $\Lambda_{\alpha\beta\theta\nu}$ is the viscosity tensor, and ${\kappa_{\alpha\beta} = (\partial_\alpha v_\beta+\partial_\beta v_\alpha)/2}$
			is the symmetrised velocity gradient tensor.
			Be aware that the relation (\ref{eqSigGITT}) is not a linear approximation of the stress-strain rate relation as the viscosity tensor is
			an, a priori very complicated, function of the shear rate $\dot\gamma$.

			In the case of our model, the quiescent contribution $\left<\sigma_{\alpha\beta}\right>^{(0)} = -p\,\delta_{\alpha\beta}$ corresponds to the
			pressure term in the modified Navier-Stokes equation.
			Let us now concentrate on the second term of Eq.~(\ref{eqSigGITT}), that corresponds to $\Delta\sigma_{\alpha\beta}$.

			From the mode coupling approximation dynamical equation, the different coefficients of the viscosity tensor can be expressed in terms
			of structural characteristics of the liquid, particularly its static structure factor ${S_q=\left<\rho_\mathbf{q}
			\rho_{-\mathbf{q}}\right>}$, and its normalised dynamical structure factor ${\Phi_q(t)=\left<\rho_\mathbf{q}(t)\rho_{-\mathbf{q}}(0)\right>
			/S_q}$.
			Defining the combinations
			\begin{equation}
				\begin{split}
					& \Sigma_\perp = T\big[S_q - S_q^2\big] \\
					& \Delta\Sigma = -T \frac{dS_{q}}{dq}\,,
				\end{split}
			\end{equation}
			and introducing the restitution coefficient $\varepsilon$, and the advected wavenumber $q_i = (\delta_{ij} + \kappa_{ij}t)q_j$,
			the viscosity tensor of an incompressible granular liquid can then be expressed in terms of five independent coefficients~:
			\begin{equation}
				\begin{split}
					& \mathcal{B}_1^P = \int_0^{+\infty}dt\int_{\mathbf{q}}\frac{1+\varepsilon}{2S_q^2}\Phi^2_{q(-t)}(t)
					\frac{q^2\Sigma_\perp\Delta\Sigma}{6q(-t)T}t \\
					& \mathcal{B}_2^P = \int_0^{+\infty}dt\int_{\mathbf{q}}\frac{1+\varepsilon}{2S_q^2}\Phi^2_{q(-t)}(t)
					\frac{q^2\Sigma_\perp\Delta\Sigma}{6q(-t)T}t^2 \\
					& \mathcal{B}_X^\sigma = \int_0^{+\infty}dt\int_{\mathbf{q}}\frac{1+\varepsilon}{2S_q^2}\Phi^2_{q(-t)}(t)
					\frac{q^3\Delta\Sigma^2}{30q(-t)T} \\
					& \mathcal{B}_1^\sigma = \int_0^{+\infty}dt\int_{\mathbf{q}}\frac{1+\varepsilon}{2S_q^2}\Phi^2_{q(-t)}(t)
					\frac{q^3\Delta\Sigma^2}{30q(-t)T}t \\
					& \mathcal{B}_2^\sigma = \int_0^{+\infty}dt\int_{\mathbf{q}}\frac{1+\varepsilon}{2S_q^2}\Phi^2_{q(-t)}(t)
					\frac{q^3\Delta\Sigma^2}{30q(-t)T}t^2 \,.
				\end{split}
			\end{equation}
			The evaluation of these coefficients generally requires a dynamical equation for $\Phi(t)$, which is very difficult to solve.
			However, in the perspective of our renormalisation group study, we can remark that $t$ and $\mathbf{q}$ are always integrated over,
			so that all the $\mathcal{B}_i^j$ are numbers that can be treated as coupling constant that get renormalised by the flow when the
			scale is changed.
			Hence, we only need the overall structure of the viscosity tensor, and not a full model to evaluate its coefficients, which greatly simplifies
			the task at hand.

			Finally, the global structure of the viscosity tensor, in the incompressible case, is given by~:
			\begin{equation}
			\label{eqTvis}
				\begin{split}
					\Lambda_{\alpha\beta\theta\nu} &= -2\mathcal{B}_1^P\delta_{\alpha\beta}\kappa_{\theta\nu} + \mathcal{B}_2^P\delta_{\alpha\beta}
					\kappa_{\theta i}\kappa_{i \nu} \\
								       &+ \mathcal{B}_X^{\sigma}X_{\alpha\beta\theta\nu} + \mathcal{B}_1^{\sigma}Y^1_{\alpha\beta\theta\nu}
								       +\mathcal{B}_2^{\sigma}Y^2_{\alpha\beta\theta\nu}\,,
				\end{split}
			\end{equation}
			where $X$ is the fully symmetrised product of Kronecker symbols~:
			\begin{equation}
				X_{\alpha\beta\theta\nu} = \delta_{\alpha\beta}\delta_{\theta\nu} + \delta_{\alpha\theta}\delta_{\beta\nu} 
				+ \delta_{\alpha\nu}\delta_{\beta\theta}\,,
			\end{equation}
			and $Y^n$ are the fully symmetric combinations of $n$ $\kappa$ tensors and Kronecker symbols~:
			\begin{equation}
				\begin{split}
					& Y^1_{\alpha\beta\theta\nu} = 2\delta_{\alpha\beta}\kappa_{\theta\nu} + \kappa_{\alpha\theta}\delta_{\beta\nu}
					+\kappa_{\beta\nu}\delta_{\alpha\theta} + \kappa_{\alpha\nu}\delta_{\beta\theta} + \kappa_{\beta\theta}\delta_{\alpha\nu} \\
					& Y^2_{\alpha\beta\theta\nu} = \delta_{\alpha\beta}\kappa_{\theta i}\kappa_{i\nu} + \kappa_{\alpha\theta}\kappa_{\beta\nu}
					+\kappa_{\alpha\nu}\kappa_{\beta\theta}\,.
				\end{split}
			\end{equation}
			This expression is robust, and allowed to recover, for example, the universal Trouton ratios in different flow geometries in the appropriate
			regimes of parameters \cite{Coquand21}.

			Our field theory is thus complete if we replace in Eq.~(\ref{eqSLambda}) $\Delta\sigma$ by its expression, ${\Delta\sigma_{\alpha\beta} =
			\Lambda_{\alpha\beta\theta\nu}\kappa_{\theta\nu}}$.
			The explicit expression of the modified Navier-Stokes equation resulting from this change are discussed in more details in terms of the
			symmetries of the equation developed in the next section.
			Note that from the above expressions, it appears that the \textit{quiescent} fluid (the one from which the background
			flow has been removed to have a better access to fluctuations) is isotropic at large scales; it should be understood however, that this
			does not imply anything on the possibility of local anisotropies that will manifest themselves in the form of vortices for example.
			The statistical average in the definition of the macroscopic stress tensor washes out these local fluctuations.

	\section{Symmetries and Ward identities}

		In this last section, we are going to discuss the symmetries of the effective average action $\Gamma_k$, without going into specific expressions
		of the model.
		These symmetries will impose constraints, known as \textit{Ward identities}, that will provide general insight on the physics of the problem,
		independently of the specific truncation scheme that can be used to perform a specific computation.
		In particular, we are going to discuss these symmetries in the light of what has already been established for the stochastic Navier-Stokes equation
		\cite{Canet15} in order to try to identify the possible sources of K41 breaking.

		\subsection{The tensor spin decomposition}

			First of all, it is instructive to decompose the velocity gradient tensor on the irreducible representations of the SO(3) group.
			We can thus write~:
			\begin{equation}
				\partial_\alpha v_\beta = \mathcal{K}^{(0)}\,\delta_{\alpha\beta} + \mathcal{K}^{(1)}_{\alpha\beta} + \mathcal{K}_{\alpha\beta}^{(2)}\,,
			\end{equation}
			where the spin 0 component $\mathcal{K}^{(0)}=$ Tr$(\kappa)$ is 0 for incompressible flows, ${\mathcal{K}^{(1)}_{\alpha\beta}=(\partial_\alpha
			v_\beta - \partial_\beta v_\alpha)/2}$ is the vorticity of the velocity field, and $\mathcal{K}^{(2)}_{\alpha\beta}$ is the traceless part of the
			symmetrised velocity gradient, which is equal to $\kappa$ in our case.

			It is not difficult to show that (i) $\Delta\sigma_{\alpha\beta}$ only involves powers of $\mathcal{K}^{(2)}$ as
			a consequence of the structure of the GITT equations, which prevents the possibilities of local breaking of equilibrium at the level of 
			individual fluid cells (note that this does not impose by any means a cancellation of the vorticity field), and (ii) that spatial
			derivatives of products of $\mathcal{K}^{(2)}$ can be expressed in terms of $\mathcal{K}^{(2)}$ only.
			Hence, it is a reasonable assumption, when wanting to keep a functional form for the stress tensor, to express it as~:
			\begin{equation}
				\Delta\sigma_{\alpha\beta}\big[\mathbf{v}\big] = \Delta\sigma_{\alpha\beta}\big[\mathcal{K}^{(2)}\big]\,.
			\end{equation}
			This replacement has consequences when examining further the symmetries of the modified Navier-Stokes equation.

			All in all, the effective average action couples the spin 0 part of the velocity gradient to the velocity response field only, the
			spin 1 part is only present in the convective term, while all the other terms (except for $\partial_t v_\alpha$) only involve the
			spin two tensor $\mathcal{K}^{(2)}$. It is thus reasonable to expect that the renormalisation of such terms can be expressed
			only in terms of $\mathcal{K}^{(2)}$.

		\subsection{Naive power counting analysis}

			A usual way to estimate whether new terms in the action are going to bring significant corrections in the renormalisation group flow is to
			perform a dimensional analysis.
			This method is called "naive" here because it is not fully rigorous~: it supposes that the correction to scaling brought by the
			renormalisation are small enough.
			Nonetheless, in absence of more information, it is usually instructive to compute the scaling dimensions of the new coupling constants.

			Plugging Eq.~(\ref{eqTvis}) into Eq.~(\ref{eqSLambda}), grouping together terms with the same number of spatial derivatives
			and powers of the velocity field, the first order correction to the stochastic Navier-Stokes equation has the form~:
			\begin{equation}
				\begin{split}
					\Delta\mathcal{S}_{vis} &= \int_x\bigg\{\frac{\mathcal{B}^P_k}{2}\overline{v}_\alpha\Big[\partial_\alpha
							\big(\partial_\beta v_\theta\big)^2
					+ \partial_\alpha\big(\partial_\beta (v_\theta)\partial_\theta(v_\beta)\big)\Big] \\
								&+\frac{\mathcal{B}^\sigma_k}{2}\overline{v}_\alpha\Big[\partial_\alpha(v_\theta)\partial^2(v_\theta)
								+ \partial_\theta(v_\alpha)\partial^2(v_\theta) \\
								&\quad+ 2\partial_\theta(\partial_\beta v_\alpha)
							\partial_\beta (v_\theta)\Big]\bigg\}
				\end{split}
			\end{equation}
			The coefficient $\mathcal{B}_X^\sigma$ corresponds to the viscous term in the SNS model, and is thus not present in the expression of
			$\Delta\mathcal{S}_{vis}$.
			Remark that since this correction is now analysed within a renormalisation group perspective, the coupling constants
			$\mathcal{B}_k$ have become dependent on the running scale $k$.
			From the form (\ref{eqTvis}) of the viscosity tensor, more terms can be generated, but we begin by the analysis of the first order
			correction in powers of the field and spatial derivatives.

			Let us define $\Delta(\cdot)$ the operator giving the scaling dimension of a given quantity, $z_k=-\Delta(t)$ the dynamical exponent,
			$\Delta\big(\mathcal{R}_k\big) = -\eta^D_k$, $\Delta\big(\nu_k\big) = -\eta^\nu_k$, and $d$ the dimension of space
			($d=3$ in our case), then, the following identities are derived~:
			\begin{equation}
			\label{eqSca}
				\begin{split}
					& z_k = 2 - \eta^\nu_k \\
					& \Delta\big(\overline{\mathbf{v}}_k\big) = \frac{d+2+\eta^D_k-\eta^\nu_k}{2} \\
					& \Delta\big(\mathbf{v}_k\big) = \frac{d-2-\eta^D_k+\eta^\nu_k}{2} \\
					& \eta^\nu_k = \frac{4-d+\eta^D_k}{3} \\
					& \Delta\big(\mathcal{B}_k^P\big) = \Delta\big(\mathcal{B}_k^\sigma\big) = \frac{-d+\eta^D_k-3\eta^\nu_k}{2}
				\end{split}
			\end{equation}
			It can be checked that these equations, except for the last one, are consistent with the usual field theory studies of the Navier-Stokes equation.
			Combining the two last equations, we get~:
			\begin{equation}
				\big(\eta^D_k\geqslant0\text{ and }\eta^\nu_k\geqslant0\big)\ \Rightarrow\ \Delta\big(\mathcal{B}_k^P\big)=\Delta\big(\mathcal{B}_k^\sigma\big)
				\leqslant0\,,
			\end{equation}
			namely, the new coupling constants are irrelevant in any dimension $d$ around the studied renormalisation group fixed point
			(negative anomalous dimensions would lead to results in contradiction with the known phenomenology of turbulent flows).

			We can even go further than that and derive the scaling relation for a coupling constant $\mathcal{B}_k^n$ corresponding to the $n$-th
			correction to the stress tensor term~:
			\begin{equation}
				\begin{split}
					\Delta\big(\mathcal{B}_k^n\big) &= \frac{(1-n)\,d-(n+1)\eta^\nu_k+(n-1)\eta^D_k}{2} \\
									&=\frac{(2-n)d-2(n+1)+(n-2)\eta_k^D}{3}\leqslant0\,.
				\end{split}
			\end{equation}
			One can check that this is consistent with Eq.~(\ref{eqSca}) when $n=2$.
			Hence, at any order, and in any dimension, the corrections to the stress tensor brings coupling constants that are all irrelevant.

			Now, let us discuss this point in more details.
			In order for the power counting analysis to hold, it must (i) be conducted close to a Gaussian fixed point, (ii) be applied in a case
			where the contribution to the renormalisation group flow are dominated by the dimensional terms, and not the non-trivial loop terms.
			None of these two conditions hold at the K41 fixed point in three dimensions.
			Therefore, it is difficult to draw solid conclusions from the above computation.
			However, it shows that if such terms are able to break the K41 scaling, then they must do so in a very non-trivial way.
			This is an important piece of information to keep in mind when designing approximations.

		\subsection{Pressure gauged shift}

			The transformation is~:
			\begin{equation}
				p(\mathbf{x},t)\mapsto p(\mathbf{x},t) + \epsilon(\mathbf{x},t)\,.
			\end{equation}
			The induced change in the action is~:
			\begin{equation}
				\delta \mathcal{S} = \int_{\mathbf{x},t}\overline{v}_\alpha(\mathbf{x},t)\partial_\alpha\epsilon(\mathbf{x},t)\,.
			\end{equation}
			Since this change is linear in the velocity response field, it can be cancelled by an appropriate redefinition of the sources.
			Writing that the generating functionals remain unchanged under this transformation leads to the following Ward identity~:
			\begin{equation}
				\left<\int_{\mathbf{x},t}\Big[\partial_\alpha\overline{v}_\alpha(\mathbf{x},t) + K(\mathbf{x},t)\Big]\epsilon(\mathbf{x},t)\right>=0
			\end{equation}
			which writes
			\begin{equation}
				\frac{\delta \Gamma_k}{\delta P} = - \partial_\alpha\overline{u}_\alpha = \Bigg<\frac{\partial \mathcal{S}}{\delta p}
				\Bigg>
			\end{equation}

			Therefore, the pressure sector is not renormalised.
			This property is inherited from the SNS model.
			It is particularly interesting to compare it with the results of the GITT model~: in \cite{Coquand20f}, it was shown that the pressure
			in a granular liquid decomposes into a term corresponding to the pressure field of an equivalent unsheared granular liquid at same
			temperature and packing fraction, that does not couple to the dynamics, and a pressure correction coming from the terms
			coupled to the dynamical structure factor $\Phi$.
			Here, the influence of the dynamical structure factor is hidden in the renormalisation of the coupling constants $\mathcal{B}_i^j$,
			but we recover the same splitting between a bare pressure, that is not coupled to the renormalisation group flow, and a pressure
			correction, generated notably by the $\mathcal{B}_k^P$ term (it generates non zero corrections on the trace of the stress tensor,
			which are nothing but pressure corrections).
			The two models, even though based on very different theoretical grounds, thus lead to comparable conclusions.
			Let us emphasize that this result is protected by the symmetries of the action, and must therefore be preserved, at any scale, by the
			renormalisation group flow; It is very firmly established.

		\subsection{Response pressure gauged shift}

			Similarly, we can study the transformation~:
			\begin{equation}
				\overline{p}(\mathbf{x},t)\mapsto \overline{p}(\mathbf{x},t) + \overline{\epsilon}(\mathbf{x},t)\,,
			\end{equation}
			which induces the following change in the action~:
			\begin{equation}
				\delta \mathcal{S} = \int_{\mathbf{x},t}\overline{\epsilon}(\mathbf{x},t)\partial_\alpha v_\alpha(\mathbf{x},t)\,.
			\end{equation}
			Again, the change is linear in the velocity field, so that it can be cancelled by an appropriate redefinition of the sources.
			Writing that the generating functionals remain unchanged under this transformation leads to the following Ward identity~:
			\begin{equation}
				\left<\int_{\mathbf{x},t}\Big[-\partial_\alpha v_\alpha(\mathbf{x},t) + \overline{K}(\mathbf{x},t)\Big]\overline{\epsilon}(\mathbf{x},t)\right>=0
			\end{equation}
			which writes
			\begin{equation}
				\frac{\delta \Gamma_k}{\delta \overline{P}} = - \partial_\alpha u_\alpha = \Bigg<\frac{\partial \mathcal{S}}{\delta \overline{p}
				}\Bigg>
			\end{equation}

			Thus, the response pressure sector is not renormalised either.
			This symmetry is crucial as it ensures that the incompressibility property is preserved at all scales.
			It is also inherited from SNS.
			Another way to phrase it is that the part of the effective action that couples to the spin 0 component of the velocity gradient field
			is not renormalised.

		\subsection{Time gauged Galilean symmetry}

			Time-gauged Galilean transformations have the form~:
			\begin{equation}
				\begin{split}
					& v_\alpha(\mathbf{x},t)\mapsto v_\alpha(\mathbf{x},t) - \frac{d\epsilon_\alpha}{dt}(t) + \epsilon_\beta(t)\partial_\beta
					v_\alpha(\mathbf{x},t) \\
					& \overline{v}_\alpha(\mathbf{x},t)\mapsto \overline{v}_\alpha(\mathbf{x},t)+\epsilon_\beta(t)\partial_\beta\overline{v}_\alpha
					(\mathbf{x},t)\\
					& p(\mathbf{x},t)\mapsto p(\mathbf{x},t) + \epsilon_\beta(t)\partial_\beta p(\mathbf{x},t) \\
					& \overline{p}(\mathbf{x},t) \mapsto \overline{p}(\mathbf{x},t) + \epsilon_\beta(t)\partial_\beta\overline{p}(x,t) \,.
				\end{split}
			\end{equation}
			After some algebra, and keeping only the first order corrections to the stress tensor, it can be shown that the change of the
			action under such a transformation is~:
			\begin{equation}
				\delta\mathcal{S} = -\int_{\mathbf{x},t}\overline{v}_\alpha\frac{d^2\epsilon_\alpha}{dt^2} = -\int_{\mathbf{x},t}\epsilon_\alpha
				\partial_t^2\overline{v}_\alpha^2 = \delta\left[\int_{\mathbf{x},t}\overline{v}_\alpha D_t v_\alpha\right]\,,
			\end{equation}
			where we introduced the covariant derivative ${D_t=\partial_t+v_\alpha\partial_\alpha}$.

			More generally, if we use the fact that $\sigma\big[\mathbf{v}\big]=\sigma\big[\mathcal{K}^{(2)}\big]$, it can be shown that all the
			terms in $\sigma$, at arbitrary order $n$, are such that $\partial_\alpha\sigma_{\alpha\beta}$ cannot bring any contribution to $\delta\mathcal{S}$.
			Hence, the tensor spin decomposition allows us to conclude that the time-gauged Galilean transformation leaves the stress tensor term
			invariant at the functional level.

			The corresponding Ward identity can be written
			\begin{equation}
				\begin{split}
					&\left<\int_{\mathbf{x},t}\epsilon_\alpha(t)\partial^2\overline{v}_\alpha(\mathbf{x},t) - \frac{d\epsilon_\alpha}{dt}(t)
					J_\alpha(\mathbf{x},t) \right. \\
					&+ \epsilon_\beta(t)\partial_\beta\big(v_\alpha(\mathbf{x},t)\big)J_\alpha(\mathbf{x},t)
					+ \epsilon_\beta(t)\partial_\beta\big(\overline{v}_\alpha(\mathbf{x},t)\big)\overline{J}_\alpha(\mathbf{x},t) \\
					&+\left.\vphantom{\int} \epsilon_\beta(t)\partial_\beta\big(p(\mathbf{x},t)\big)K(\mathbf{x},t)
						+\epsilon_\beta(t)\partial_\beta\big(\overline{p}(\mathbf{x},t)\big)\overline{K}(\mathbf{x},t) \right>=0
				\end{split}
			\end{equation}
			or equivalently,
			\begin{equation}
				\begin{split}
					& \int_{\mathbf{x},t}\bigg\{\partial_t^2\overline{u}_\alpha + \partial_t\left(\frac{\delta\Gamma_k}{\delta u_\alpha}\right)
						+\partial_\alpha\big(u_\beta\big)\frac{\delta \Gamma_k}{\delta u_\beta}
						+\partial_\alpha\big(\overline{u}_\beta\big)\frac{\delta\Gamma_k}{\delta \overline{u}_\beta} \\
					& +\partial_\alpha\big(\overline{P}\big)\frac{\delta\Gamma_k}{\delta \overline{P}} + \partial_\alpha
				\big(P\big)\frac{\delta \Gamma_k}{\delta P}\bigg\} = 0\,.
				\end{split}
			\end{equation}
			Hence, for the first term, that can be written ${\displaystyle\int_{\mathbf{x},t}\overline{u}_\alpha(\mathbf{x},t)
			D_t u_\alpha(\mathbf{x},t)}$,
			the variation of the effective average action is the same as that of the microscopic action $\mathcal{S}$.
			$\Gamma_k$ can thus be written as the sum of a covariant derivative term, that does not get renormalised, and a Galilean invariant part.

			Before going further, let us pause for a moment to discuss this statement.
			The non-renormalisation theorem of the covariant derivative part of the coarse grained velocity field is to be compared with the
			work of Cartes and al. \cite{Cartes22}.
			In this article, a study of the inviscid limit of the unforced Burgers equation is presented.
			It is shown that under these condition, a steady solution corresponding to $D_t u=0$ exists, namely, in the inviscid limit, the time
			derivative and advection term can exchange energy without breaking the out of equilibrium steady state, said otherwise,
			there exists stable flows with only time derivative and advection competition.
			As a result, the existence of a non-trivial energy cascade requires the existence of a non negligible stochastic forcing, and
			a response from the fluid (embodied here by the stress tensor).
			This property must certainly hold also in our case in the limit where the stress tensor contributions go to zero, up to the fact that the
			SNS velocity field is divergenceless, while the Burgers one is solenoidal.

			Next, let us examine the above statement in the light of the tensor spin decomposition.
			As explained above, the only term that couples to $\mathcal{K}^{(1)}$, or the vorticity, is the advection term.
			The fact that the covariant derivative is not renormalised means that the renormalisation of the spin 1 component of the velocity
			gradient is entirely fixed by the renormalisation of the spin 2 part, and the velocity field itself --- the $\partial_t u$ term is not
			expressed in terms of velocity gradients.
			This property holds both in 2 and in 3 dimensions, which means that, even in two dimensions, the renormalisation of the enstrophy
			is entirely determined by the renormalisation of the symmetrised velocity gradient, and the time derivative of the velocity field.
			Be aware that this is not a statement about the simplicity or complexity of the expression of such a flow.

		\subsection{Time-gauged response fields shift symmetry}

			This transformation has the following form~:
			\begin{equation}
				\begin{split}
					& \overline{v}_\alpha(\mathbf{x},t)\mapsto\overline{v}_\alpha(\mathbf{x},t) + \overline{\epsilon}_\alpha(t) \\
					& \overline{p}(\mathbf{x},t)\mapsto \overline{p}(\mathbf{x},t) + v_\beta({\mathbf{x},t})\overline{\epsilon}_\beta(t)\,.
				\end{split}
			\end{equation}

			The variation of the action is once again linear in the fields~:
			\begin{equation}
				\delta\mathcal{S} = \int_{\mathbf{x},t}\left\{\overline{\epsilon}_\beta(t)\partial_tv_\beta(\mathbf{x},t)
				+2\overline{\epsilon}_\alpha(t)N_{\alpha\beta}(x)\overline{v}_\beta(\mathbf{x},t)\right\}\,,
			\end{equation}
			which leads to the Ward identity~:
			\begin{equation}
				\begin{split}
					\int_{\mathbf{x}}\partial_t u_\beta(\mathbf{x}) &+ 2\left<\int_{\mathbf{x}}N_{\beta\gamma}(x)
					\overline{v}_\gamma(\mathbf{x})\right>\\
											     &=\int_{\mathbf{x},t}\left\{\frac{\delta\Gamma_k}{\delta \overline{u}_\beta}
											     +u_\beta\frac{\delta\Gamma_k}{\delta \overline{P}}\right\}\,.
				\end{split}
			\end{equation}
			Again, the right-hand side of this equation is similar to the effect of the transformation on the microscopic action, which means that the left-hand
			side is not renormalised.

			Putting together all the results we have so far, we can thus write that~:
			\begin{equation}
				\begin{split}
					&\Gamma_k\big[\mathbf{u},\overline{\mathbf{u}},P,\overline{P}\big] = \int_{\mathbf{x},t}\overline{u}_\alpha(\mathbf{x},t)
					\big\{D_t u_\alpha(\mathbf{x},t)+ \partial_\alpha P(\mathbf{x},t)\big\} \\
					&+ \int_{\mathbf{x},t}\overline{P}(\mathbf{x},t) \partial_\alpha u_\alpha(\mathbf{x},t) \\
					&  - \int_{\mathbf{x},\mathbf{x'},t}\overline{u}_\alpha(\mathbf{x},t) N_{\alpha\beta}(|\mathbf{x}-\mathbf{x'}|)
					\overline{u}_\beta(\mathbf{x'},t)
					 +\tilde{\Gamma}_k[\mathbf{u},\overline{\mathbf{u}}]\,,
				\end{split}
			\end{equation}
			where $\tilde{\Gamma}_k$ is invariant under both Galilean and response field shift transformations.

			It is remarkable that this result is exactly the same as the one established by the group of L. Canet in the case of the SNS model ten
			years ago \cite{Canet15}.
			This is a strong result~: The dynamical equations for the steady flowing state of granular liquids shares almost all the symmetries of the
			SNS model.
			In particular, it has the same behavior under the time-gauged Galilean symmetries.
			The breaking of K41 is thus realised in a quite subtle way.
			Unlike in the case of the qualitative arguments presented in the first section, we can see that it is not so easy to find symmetries of the SNS
			equation that are broken by the granular liquids.

		\subsection{Time- and space-gauged shift of the response field}

			In this last paragraph, we examine the effect of the following transformation~:
			\begin{equation}
				\begin{split}
					& \overline{v}_\alpha(\mathbf{x},t) \mapsto \overline{v}_\alpha(\mathbf{x},t) + \overline{\epsilon}_{\alpha}(\mathbf{x},t) \\
					& \overline{p}(\mathbf{x},t) \mapsto \overline{p}(\mathbf{x},t) + v_\beta(\mathbf{x},t)\overline{\epsilon}_\beta(\mathbf{x},t)\,,
				\end{split}
			\end{equation}
			but in addition, following \cite{Canet15}, we add a new source term, $L$, that couples to the composite operator of the squared
			velocity as follows $v\cdot L\cdot v$.
			Once again, the induced transformation of the microscopic action is linear in the fields, so that the effect of the transformation above
			can be reabsorbed by a shift in the sources,leading to the following Ward identity~:
			\begin{equation}
				\begin{split}
					& \left<-\partial_t v_\alpha -\partial_\alpha p+\partial_\beta\Delta\sigma_{\alpha\beta}\big[\mathcal{K}^{(2)}\big]
					-\partial_{\beta}\big(v_\alpha v_\beta\big) \right. \\
					& \left.+\overline{J}_\alpha + \overline{K}v_\alpha+ 2N_{\alpha\beta}*\overline{v}_\beta\vphantom{\Big]}\right>= 0 \,,
				\end{split}
			\end{equation}
			where $*$ is the convolution operator.
			Still following \cite{Canet15}, we can reexpress this equation in terms of the generator of connected correlation functions $\mathcal{W}_k$~:
			\begin{equation}
				\begin{split}
					& -\partial_t\frac{\delta \mathcal{W}_k}{\delta J_\alpha} - \partial_\alpha\frac{\delta \mathcal{W}_k}{\delta K}
					+\left<\partial_\beta\Delta\sigma_{\alpha\beta}\left[\left\{\left(\frac{\delta\mathcal{W}_k}{\delta J_\alpha}\right)^n
					\right\}_{n\in\mathbb{N}}\right]\right> \\
					& +\overline{J}_\alpha + \overline{K}\frac{\delta\mathcal{W}_k}{\delta J_\alpha}-\partial_\beta\frac{\delta\mathcal{W}_k}
					{\delta  L_{\alpha\beta}} + 2N_{\alpha\beta}*\frac{\delta\mathcal{W}_k}{\delta\overline{J}_\beta} = 0\,.
				\end{split}
			\end{equation}
			Differentiating the result with respect to the source $J_\alpha$, and taking the limit of zero sources allows to reexpress the derivatives
			of $\mathcal{W}_k$ in terms of the fields expectation values~:
			\begin{equation}
				\begin{split}
					& -\partial_t\left<v_\alpha(\mathbf{x},t)v_\alpha(\mathbf{y},t)\right> - \partial_\gamma^x\left<v_\alpha(\mathbf{x},t)
					v_\gamma(\mathbf{x},t)v_\alpha(\mathbf{y},t)\right> \\
					&- \partial_\gamma^y\left<v_\alpha(\mathbf{y},t)v_\gamma(\mathbf{y},t)v_\alpha(\mathbf{x},t)\right>
					+\left<f_\alpha(\mathbf{x},t)v_\alpha(\mathbf{y},t)\right> \\
					&+\left<f_\alpha(\mathbf{y},t)v_\alpha(\mathbf{x},t)\right> \\
					&+\left(\frac{\delta}{\delta J_\alpha}\partial_\gamma
						\Delta\sigma_{\gamma\alpha}\left[\left\{\left(\frac{\delta \mathcal{W}_k}{\delta J_\theta}\right)^n\right\}_{n\in\mathbb{N}}
					\right]\right)_{U=0} = 0\,.
				\end{split}
			\end{equation}
			In the case where the stress tensor contribution reduces to the viscous term, this equation is nothing but the Kármán-Howarth relation
			\cite{Canet15}, which yields the Kolmogorov 4/5th law, and therefore the K41 scaling in the energy cascade
			(note that although this is not the usual derivation of the 4/5th law, it is strictly equivalent to it).
			But in the case of granular liquids, as we have shown above, more contributions are to be expected.
			In particular, terms of the for ${\displaystyle \partial_\theta\big(\partial_\alpha v_\beta\big)^2}$ will give new terms
			of the form ${\displaystyle \partial_\theta\left<\partial_\alpha\big(v_\beta v_\gamma\big)\partial_\alpha v_\beta\right>}$
			that have no good reason to cancel a priori.

			All in all, even though the naive power counting could have lead us to think that the new contributions to the stress tensor, being
			a priori irrelevant, would play a negligible role, we have just shown that they break explicitly the symmetry leading to the
			Kármán-Howarth relation in the SNS model, namely to the K41 scaling.

			When studying the renormalisation group flow of the modified SNS model presented in this paper, it is expected that the usual SNS results
			are recovered if the initial condition at the microscopic scale is ${\forall i,j\:\mathcal{B}_i^j=0}$.
			This result is consistent with the known rheology of granular liquids, that permits a Newtonian regime when the driving power is strong
			and the packing fraction is not too high \cite{Kranz18}.
			However, we expect this submanifold to be unstable with respect to the addition of non trivial higher order corrections to the stress tensor
			term.
			What remains an open question is if the proper universality class can be generated with the few first order corrections to the stress tensor,
			or if a more global treatment is needed.
			We reserve the answer to this question to future work.

\section{Conclusions}

	In conclusion, we have provided general guidelines to build a field theory model that should enable to study the energy cascade in granular liquids
	in steady shear flows.
	This model combines both elements from the field theory study of dynamical systems, and granular specific material borrowed to the GITT model.

	Our main objective was to study the possible breaking of K41 in such systems.
	As it turned out, even if qualitatively, the numerical result of Saitoh et al. \cite{Oyama19} was easily recovered, we have shown more
	rigorously that most of the symmetries protecting the establishment of scaling relations in the SNS model are shared in the granular case.
	Only one of them is broken explicitly by the new contributions to the stress tensor.
	As it turns out, this symmetry is precisely the one that lead to the Kármán-Howarth relation, that is, to the K41 scaling.
	We can thus expect a new universality class to arise in those systems.

	At this stage, we have opened the possibility of existence of a new scaling Universality class, but we did not predict any value for the related exponents,
	nor did we conclude on the strength of intermittency effects in such systems.
	Answering these questions requires to build an explicit approximation scheme on the Wetterich equation (\ref{eqWe}) and will be examined in following studies.

\ack{
	The work on GITT used in this article owes a lot to T. Kranz and M. Sperl.
	I am very grateful to L. Canet and L. Gosteva for their great patience and useful discussions.
	I also would like to thank P. Rognon and F. Radjai that very nicely accepted to share their knowledge of simulations of granular systems,
	and A. Zaccone for his constructive comments about the manuscript.
}

	\printbibliography

\end{document}